\documentclass[a4paper]{article}

\usepackage{amssymb}
\usepackage{amsmath}

\usepackage{xcolor}
\usepackage{mathrsfs}
\usepackage{booktabs}
\usepackage{graphicx}
\usepackage[authoryear,round]{natbib}
\usepackage{subfig}
\usepackage{mathrsfs}
\usepackage{fullpage}
\usepackage[hidelinks]{hyperref}
\begin{document}

\global\long\def\vect#1{\boldsymbol{\mathrm{#1}}}%

\global\long\def\Div{\mathrm{Div\,}}%

\global\long\def\diver{\mathrm{div\,}}%

\global\long\def\sym#1{\mathrm{Sym}#1}%

\global\long\def\skw#1{\mathrm{Skw}#1}%

\global\long\def\psym#1{\mathrm{PSym}#1}%

\global\long\def\lin#1{\mathrm{Lin}#1}%

\global\long\def\orth{\mathrm{Orth}^{+}}%

\global\long\def\tr#1{\mathrm{tr}#1}%

\global\long\def\d{\mathrm{d}}%

\global\long\def\phy{\varphi}%

\global\long\def\com#1{\textcolor{red}{#1}}%

\global\long\def\epsilon{\varepsilon}%

\global\long\def\defeq{\vcentcolon=}%

\global\long\def\ring#1{\mathring{#1}}%

\global\long\def\red#1{{\color{red}#1}}%

\global\long\def\blue#1{{\color{blue}#1}}%

\global\long\def\grad#1{\mathrm{grad}(#1)}%

\global\long\def\argmin#1#2{\underset{#2}{\mathrm{argmin\,}}\left\{  #1\right\}  }%

\global\long\def\argmax#1#2{\underset{#2}{\mathrm{argmax\,}}\left\{  #1\right\}  }%

\global\long\def\tens#1{\mathsf{#1}}%

\global\long\def\Diver{\Div}%

\title{\textsc{Elastocapillary morphing of self-encapsulated  droplets floating at the oil-air interface}}

\author{\textsc{D. Andrini}$^1$ $\,\,\cdot$
    \textsc{D. Riccobelli}$^2$ $\,\,\cdot$
    \textsc{L. Gazzera}$^3$ \\
    \textsc{S. Molteni}$^3$ $\,\,\cdot$
    \textsc{P. Metrangolo}$^3$ $\,\,\cdot$
    \textsc{P. Ciarletta}$^4$\\
    \normalsize $^1$DISMA, Politecnico di Torino, Corso Duca degli Abruzzi, 24, Torino, Italy\\
    \normalsize Department of Engineering Mechanics, and Soft Matter Research Center,\\
    \normalsize Zhejiang University, Hangzhou 310027, P.R. China.\\
    \normalsize $^2$MathLab, Mathematics Area, SISSA -- International School for Advanced Studies,\\
    \normalsize Via Bonomea 265, Trieste, Italy.\\
    \normalsize $^3$Department of Chemistry, Materials, and Chemical Engineering ``Giulio Natta'',\\
    \normalsize Politecnico di Milano, Piazza Leonardo da Vinci 32, Milano, Italy.\\
    \normalsize $^4$MOX, Department of Mathematics, Politecnico di Milano,\\
    \normalsize Piazza Leonardo da Vinci 32, Milano, Italy.}
\date{\today}

\maketitle
\begin{abstract}
Self-encapsulated droplets floating at an oil--air interface undergo striking shape changes during evaporation, including flattening and localized loss of membrane tension leading to crumpling and wrinkling. Here we combine experiments, modeling and simulations to obtain predictive morphological maps. We perform contact-angle and evaporation experiments on water droplets coated by a hydrophobin protein film and floating in a fluorinated oil, providing reference profiles and volume-loss sequences for quantitative validation. We develop an axisymmetric mechanics framework in which equilibria follow from minimization of a total free energy combining surface energies, membrane strain energy and gravitational potential, subject to volume and contact-line constraints. A quasi-convex tension-relaxation rule accounts for compression-free states and enables coexistence of taut, wrinkled (one principal tension vanishes) and crumpled (both vanish) membrane domains. A finite element algorithm computes quasi-static morphing under volume reduction; key parameters are identified by fitting the reference contact-angle profile and then used without further tuning. The model reproduces the experimentally observed shape evolution and resolves the associated stress redistribution. Systematic parameter scans yield morphological phase diagrams governed by the Bond number, the oil--droplet surface-tension ratio and the density ratio. For buoyant droplets, crumpling relocates between exposed and submerged caps as parameters vary; for heavy droplets, a crossover to circumferential wrinkling along the immersed sidewall emerges. Wall-meniscus variations shift phase boundaries and can suppress bottom crumpling, consistent with wall-affected experiments.
\end{abstract}

\section{Introduction}
Tiny liquid droplets can float on top of other immiscible liquids thanks to the interplay of capillarity and gravity \citep{langmuir1933oil,de1985wetting, burton2010experimental,wong2017non}. Examples include hydrocarbon droplets on water, oil-water interactions in cooking, and certain pharmaceutical formulations involving lipid-based droplets in aqueous media. The encapsulation of liquid droplets \citep{lathia2023tunable} has been exploited in both culinary and pharmaceutical contexts. In molecular gastronomy, for example, flavored liquid droplets can be coated with a thin membrane to create encapsulated spheres \citep{fu2014material, mou2020controllable}, while in pharmaceuticals, lipid-based droplets can encapsulate drugs to improve stability, solubility, and targeted delivery. Self-encapsulating Poly(lactic-co-glycolic acid) (PLGA) microspheres are being explored, for example, as a means to reduce or eliminate multiple booster injections in vaccine delivery \citep{desai2013active,reinhold2012self,mazzara2019self}

By controlling the shape of the droplets during encapsulation, it is possible to tailor fluid–structure interactions in flowing systems, such as drug delivery, or to create specific sensory effects in foods. In culinary applications, encapsulated droplets can deliver concentrated flavors, aromas, or textures in a controlled manner, allowing chefs to design multi-sensory experiences \citep{given2009encapsulation}. The size, shape, and mechanical properties of the droplets influence how they burst, release their contents, and interact with other components on the palate, making droplet encapsulation a powerful tool in molecular gastronomy. In engineering applications, the shape control of encapsulated droplets can be driven by elasto-capillary interactions with an external medium \citep{py2007capillary,prasath2021wetting}, or by the orientation of the droplets with respect to the applied external forces, such as gravity \citep{abkarian2013gravity}. 
Another mechanism is evaporation of colloidal droplets, thanks to which
particles are brought to the surface during fluid movement and self-assemble at the boundary \citep{li2024evaporative}. This mechanism leads to a variety of phenomena, including the so-called coffee-ring effect \citep{deegan1997capillary} to self-encapsulation of droplets.

The evaporation of encapsulated liquid droplets can lead to different morphologies due to fluid-structure interaction and structural buckling of the surrounding shell \citep{pauchard2003mechanical,tsapis2005onset,bala2005non,wulsten2009levitated,basu2016towards}. 

Recent studies have shown that the combined action of gravity and elasto-capillarity can produce a rich variety of morphologies in evaporating droplets resting on rigid substrates. In particular, droplets containing water and hydrophobin (HFBII), a protein obtained from \textit{Trichoderma} fungi, can spontaneously form an elastic coating once the protein concentration reaches a critical threshold \citep{yamasaki2016flattened,yamasaki2016formation,riccobelli2023flattened}. Depending on the orientation of the droplet with respect to gravity, the resulting morphologies may exhibit flattened regions or circumferential wrinkling \citep{yamasaki2016flattened,yamasaki2016formation,knoche2013elastometry,riccobelli2023flattened}. In particular, ~\cite{riccobelli2023flattened} showed that, for droplets evaporating on a rigid solid substrate, these shape changes are governed by the elastocapillarity of the interfacial film and gravity. Moreover, flattened regions are associated with coating crumpling caused by the release of compressive elastic energy. 

The present paper addresses a different physical problem. Here, the encapsulated droplet rests on an immiscible liquid substrate rather than on a rigid solid support. This change qualitatively modifies the mechanics of the system because the equilibrium shape is now determined by the coupled deformation of the droplet and of the underlying liquid substrate, together with the competition among buoyancy, capillarity, and compression of the interfacial film during evaporation. Compared to Riccobelli et al. (2023), the present work further introduces three main novelties. 
On the experimental side, we investigate evaporation-driven morphological transitions and systematically assess the role of wall-meniscus interactions. 
On the theoretical side, we incorporate a relaxed energy density that captures tension relaxation and allows for the emergence of wrinkled and crumpled regions. 
On the numerical side, we construct phase diagrams that quantitatively predict the observed transitions as functions of control parameters.

The paper is organized as follows. In Section~2, we present the experimental setup and summarize the main observations. The mathematical model of the self-coating floating droplet is derived in Section~3. A numerical method to integrate the resulting boundary-value problem is presented in Section~4, while the results, together with a comparison with experimental observations, are discussed in Section~5. Finally, concluding remarks and potential future developments are presented in Section~6.

\section{Material and Methods}
\noindent We summarize the materials and experimental methods used in this work.

\subsection{Physical system and experimental rationale}

\begin{figure}[h!]
\centering
\includegraphics[width=0.9\textwidth]{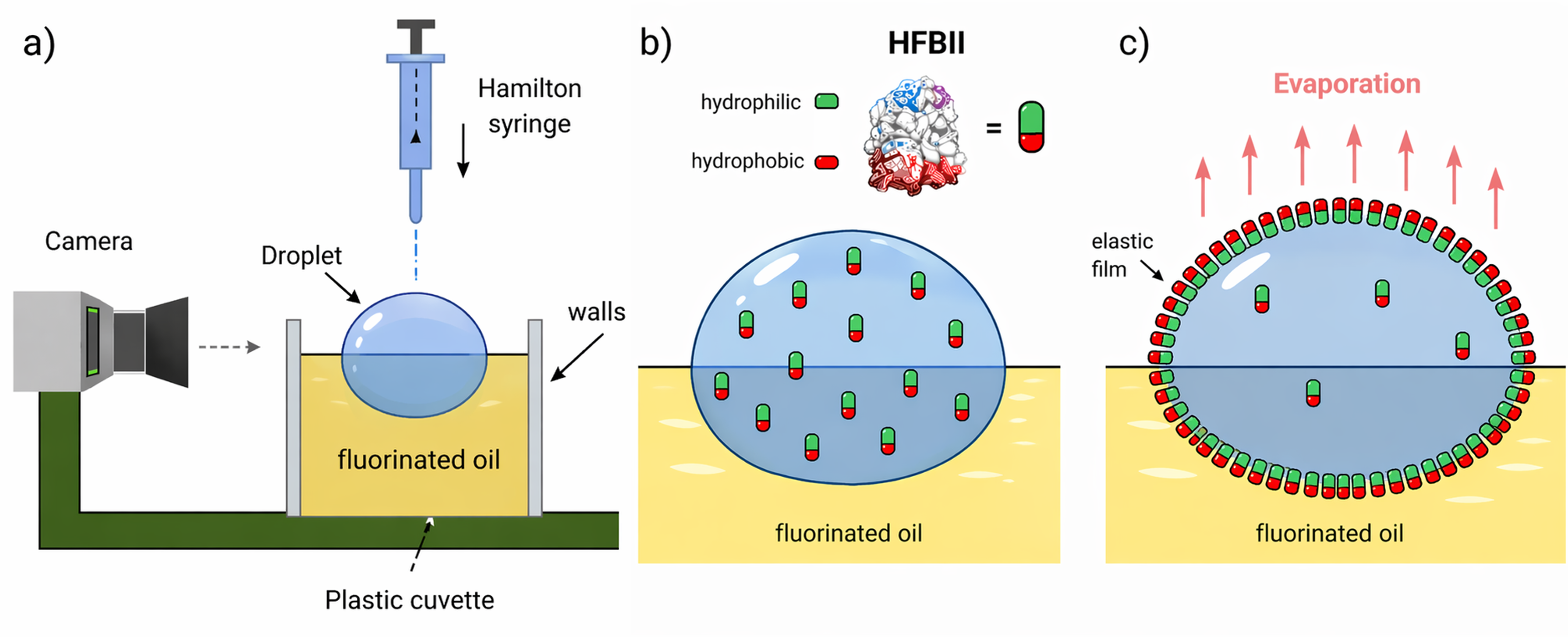}
\caption{Schematic illustration of the experimental setup and of the physical mechanism underlying the observed morphologies. 
(a) A droplet of HFBII aqueous solution is deposited on a fluorinated oil bath inside a plastic cuvette and imaged by a side-view camera. 
(b) HFBII molecules are initially dispersed within the droplet and progressively adsorb at the interfaces owing to their amphiphilic nature. 
(c) During evaporation, the increase in HFBII concentration leads to the formation of an interfacial elastic film.}
\label{fig0}
\end{figure}

We investigate the evaporation of aqueous droplets containing hydrophobin HFBII while floating on a denser fluorinated oil. As sketched in Figure \ref{fig0}, this configuration involves a three-phase system (air--water--oil) in which interfacial tension, gravity, and evaporation-induced compositional changes are strongly coupled.

As evaporation proceeds, solvent loss leads to an increase in HFBII concentration within the droplet. Because of its amphibolic nature, HFBII progressively adsorbs at the air--water and oil--water interfaces, forming an interfacial layer that evolves toward an elastic membrane. This process introduces a non-trivial coupling between capillarity and interfacial mechanics, which is central to the phenomena studied in this work.

The experiments are designed to address two main objectives. First, we characterize the morphological evolution during evaporation in a reference configuration where boundary-induced meniscus effects are minimized. Second, we assess the influence of wall-induced capillary interactions by modifying the external meniscus through controlled changes in the filling condition of the container.

The relevant observables for the present study are:
(i) the droplet profile during evaporation,
(ii) the shape evolution of the submerged and exposed caps,
(iii) the emergence of a flattened or crumpled region at the bottom interface, and
(iv) the modification of this morphological pathway in the presence of wall meniscus effects.

\subsection{Contact angle experiments}

Contact angle experiments were performed using the OCA 15 plus apparatus (Dataphysics).  In the reference configuration, a $5\,\mu\mathrm{L}$ droplet of HFBII solution  (initial concentration $0.1\,\mathrm{mg/mL}$)  is deposited on a fluorinated oil bath contained in a cuvette filled to the brim. This configuration minimizes the visible outer meniscus and provides a nearly axisymmetric floating droplet.

Side-view images are acquired at a rate of one frame per minute. The resulting sequence reveals a well-defined morphological evolution during evaporation, as shown in Figure \ref{exp1}. At early times, the droplet shape remains close to its initial configuration, with only minor variations in curvature. As evaporation progresses, the lower (oil-facing) region undergoes a progressive loss of curvature and develops a flattened segment. This region first assumes a trapezoidal shape and subsequently approaches a nearly rectangular profile. In contrast, the upper (air-facing) cap remains comparatively less affected.

This asymmetric deformation indicates that volume loss is not accommodated uniformly along the interface. Instead, it suggests a progressive redistribution of interfacial stresses, with localized relaxation in the submerged region, consistent with the formation of a mechanically active interfacial layer.

\subsection{Effect of wall meniscus interaction}

To investigate the role of boundary-induced capillary effects, a second set of experiments is performed in a cuvette filled only partially, so that a pronounced oil meniscus develops at the wall. The droplet is released in close proximity to the wall, thereby introducing a controlled interaction with the external meniscus.

Under these conditions, the morphological evolution differs qualitatively from the reference case, as shown in Figure \ref{exp2}. In particular, the flattening of the lower interface is significantly reduced or suppressed, and the overall droplet shape remains more rounded throughout the evaporation process.

These observations indicate that the external meniscus alters the capillary loading transmitted to the droplet and modifies the balance of interfacial stresses. The wall-induced meniscus therefore acts as an additional control parameter governing the selection of the droplet morphology during evaporation.

\subsection{Materials and experimental procedures}

Class II hydrophobin HFBII produced from recombinant
strains of T. reesei was kindly provided by VTT Technical Research Centre
of Finland. It was purified and lyophilized by RP-HPLC and stored at room temperature under vacuum. The purified protein is dissolved in Milli-Q water at a concentration of $0.1\,\mathrm{mg/mL}$.

Droplets of volume $5\,\mu\mathrm{L}$ are generated using a Hamilton gas-tight syringe and gently released onto the oil surface. The oil phase consists of Fomblin Y LVAC 25/6, whose density exceeds that of water, allowing the droplet to float with a partially immersed configuration.

Experiments are conducted in plastic UV cuvettes. Two filling conditions are considered: (i) cuvettes filled to the brim, to minimize the visible wall meniscus, and (ii) partially filled cuvettes, to generate a pronounced meniscus and induce droplet--wall interaction. In both cases, droplet deposition is achieved by a controlled up-and-down motion of the syringe until full detachment is obtained.
Images are recorded using an OCA 15 plus instrument equipped with a side-view camera, with a temporal resolution of one image per minute.

\begin{figure}[ht]
\centering
\includegraphics[width=0.9\textwidth]{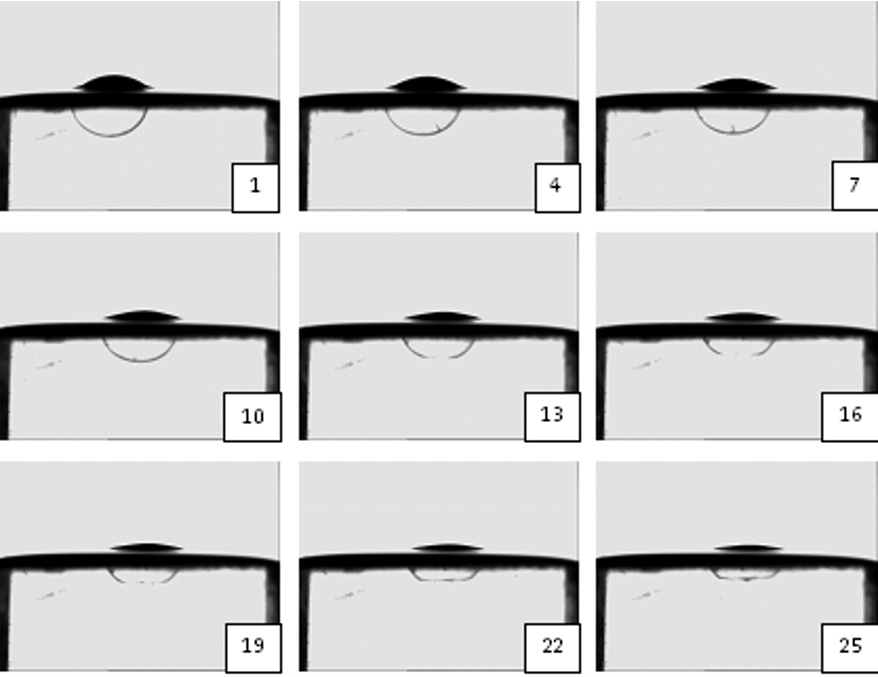}\caption{Experimental images of the evaporation of a $5\,\mu L$ droplet of HFBII on Fomblin Y LVAC 25/6. Labels indicate the time after deposition in minutes.}
\label{exp1}
\end{figure}

\begin{figure}[h!]
\centering
\includegraphics[width=0.99\textwidth]{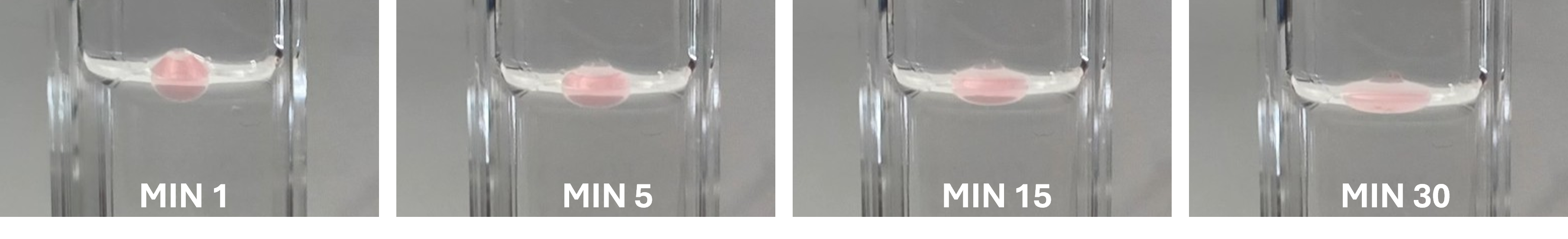}
\caption{Experimental images of the evaporation of a $5\,\mu L$ droplet of HFBII on Fomblin Y LVAC 25/6 in a partially filled cuvette, highlighting the effect of wall-meniscus interactions. Labels indicate the time after deposition in minutes.}
\label{exp2}
\end{figure}

\subsection{Image processing and extraction of droplet profiles}

The acquired images are processed using standard image analysis techniques. 
 First, the pictures are converted to binary images using ImageJ and processed using the library MorphoLibJ \citep{legland2016morpholibj}.
The droplet interface is determined by extracting the boundaries using the Python library OpenCV. Finally, the
drop profiles obtained in this way are processed using Mathematica (Wolfram Research, Inc., Version 14.3, Champaign, IL, USA), to extract information such as the volume of
the drop, the diameter of the flat spot of the surface, and the height of the drop. 

Each frame is binarized to extract the droplet contour, which is then used to reconstruct the two-dimensional profile of the droplet.

The extracted profiles serve as reference data for the subsequent analysis. In particular, they are used for parameter identification, for comparison with theoretical predictions, and for characterizing the morphological transitions observed during evaporation.

\section{Mathematical Model}
We consider an isolated axisymmetric droplet composed of water and HFBII floating in a finitely extended container filled with Fomblin Y. As water slowly evaporates, the HFBII concentration increases until it reaches a critical threshold, triggering the rapid formation of an elastic membrane that encapsulates the droplet.

Since evaporation occurs on a much slower timescale than the droplet's internal dynamics, we treat the process as quasi-static and assume mechanical equilibrium at all times.
We also treat the encapsulation process as instantaneous as it is in practice much faster than evaporation.

To solve the elastic problem of the encapsulated droplet, we need a reference configuration. We define it as the state that occurs when the HFBII concentration reaches its critical value and the membrane forms.
In the following, we will detail the mathematical model to determine both the reference configuration and the subsequent deformed one, which we refer to as the current configuration, resulting from further water evaporation beyond the critical volume. 
\subsection{Kinematics and geometric assumptions}

We characterize the state of the droplet through the interfaces that separate the three phases involved, namely water, oil, and air. 
Specifically we denote with ${\cal M}_0$ and $\cal M$ the collection of the interfaces in the reference and current configurations, respectively.

\begin{figure}
\begin{centering}
\includegraphics[width=0.9\textwidth]{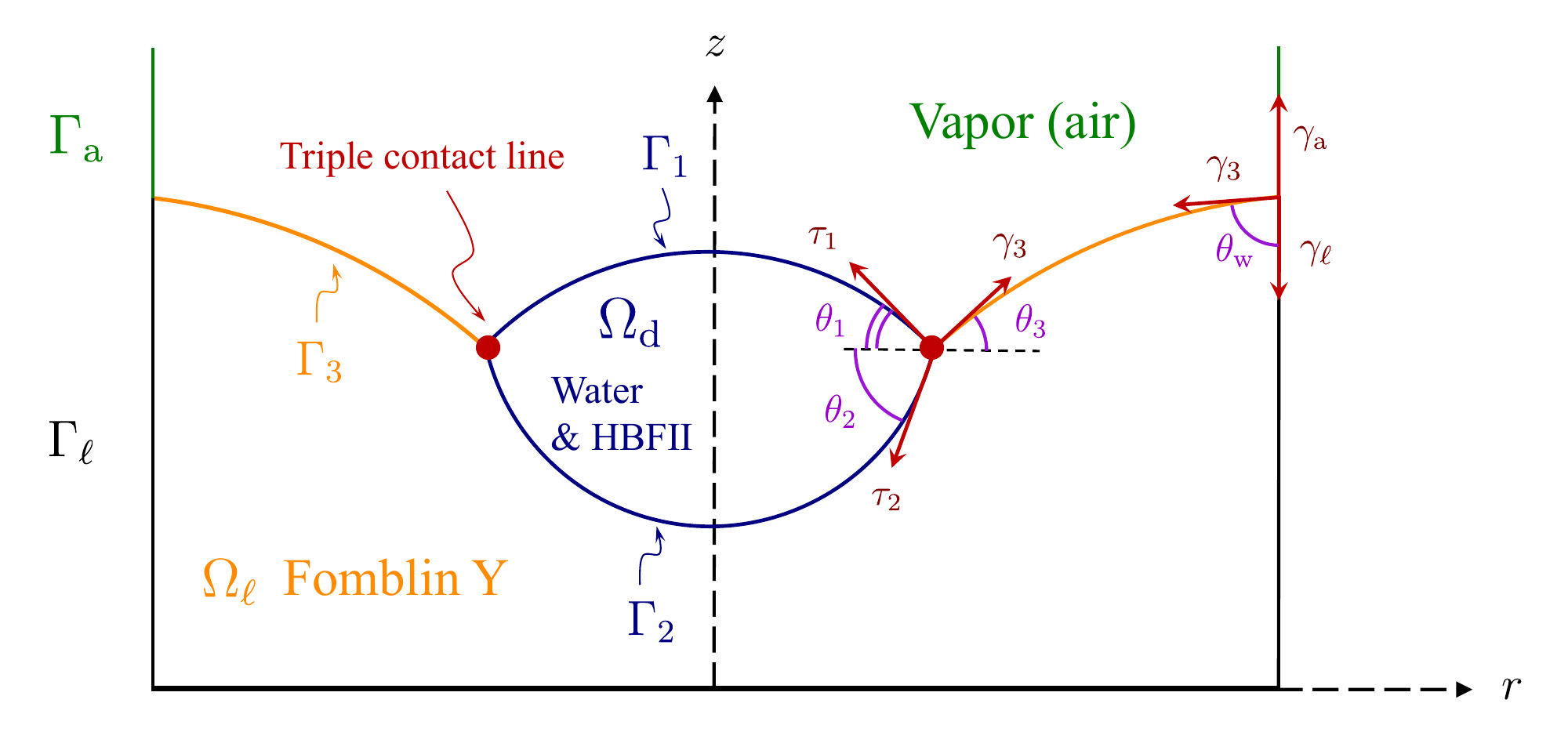}\caption{Section view of the droplet. Highlighted in blue the profiles $\Gamma_1$ and $\Gamma_3$ enclosing the droplet whose domain is denoted with $\Omega_\d$. In yellow we depict the outer meniscus $\Gamma_3$ delimiting the region occupied by the oil that we denoted with $\Omega_\ell$. 
The profiles $\Gamma_i$ meet at at the {triple contact line} that we draw with a red dot.
The black lines delimit container's wall in touch with oil ($\Gamma_\ell$) while in green the ones in touch with air ($\mathrm{\Gamma_\mathrm{a}}$).
The red arrows depict the membrane tensions at the triple contact line and at the walls, together with their corresponding angles with the floor. This illustration refers to the {current} configuration of the droplet. The one for the {reference} configuration is completely analogous except for the tensions $\tau_1$ and $\tau_2$ at the contact line that are replaced with the surface tensions $\gamma_1$ and $\gamma_2$.
\label{fig:sketch-droplet}}
\par\end{centering}
\end{figure}

Both ${\cal M}_0$ and ${\cal M}$ are surface manifolds embedded in the Euclidean space $\mathbb{E}^3$.
As shown in Fig.~\ref{fig:sketch-droplet}, we assume that ${\cal M}$ is the union of three connected surfaces,
$\Gamma_{1}$, $\Gamma_{2}$, and $\Gamma_{3}$.
Specifically, $\Gamma_{1}$ is the droplet--air interface (i.e., the upper part of the droplet),
$\Gamma_{2}$ is the droplet--oil interface, and $\Gamma_{3}$ is the oil--air interface, namely the outer meniscus.
The three surfaces meet along the triple contact line, while the outer meniscus remains in contact with the container wall.
The same assumptions apply to the reference configuration, ${\cal M}_0=\bigcup_{i}\Gamma_{0i}$, where $\Gamma_{0i}$
denotes the counterpart of $\Gamma_i$ in the reference configuration.
We refer to manifolds with these features as {admissible profiles}.
Finally, we denote by $\Gamma_\ell$ the portion of the container surface in contact with the oil, and by
$\Gamma_\mathrm{a}$ the portion in contact with the vapor.

We keep track of elastic deformations of the droplet film by means of a deformation map $\vect{f}_{\alpha}: \Gamma_{0\alpha} \to \Gamma_{\alpha}$ that maps material points belonging to $\Gamma_{0\alpha}$ into spatial points in $\Gamma_{\alpha}$, with $\alpha=1,2$.
Throughout the rest of the paper, latin indices range in $\{1,2,3\}$ and greek ones range in $\{1,2\}$.
We denote by $\tens{F}_{\alpha}$ the in-plane deformation gradient and with $\tens{C}_\alpha = \tens{F}_\alpha^T \tens{F}_\alpha$ the in-plane right Cauchy-Green strain tensor.

\subsubsection{Axisymmetric droplet}

In the following, we assume that the droplet is axisymmetric, floating in a cylindrical container of radius $R$. Let us then endow $\mathbb{E}^3$ with cylindrical coordinates $(r,\,\theta,\,z)$ and denote by $\vect{e}_r,\vect{e}_\theta,\vect{e}_z$ the corresponding coordinate vectors.

Thanks to axisymmetry, the profiles $\Gamma_{0i}$ and $\Gamma_i$ can be obtained by revolving planar curves around the $z$ axis. We parametrize them as%
\begin{equation}
     \left\{
     \begin{aligned}
         &\vect{\Gamma}_{0i}(s_{i},\theta):=r_{0i}(s_{i})\vect e_{r}(\theta)+z_{0i}(s_{i})\vect e_{z},\\
         &\vect{\Gamma}_{i}(s_{i},\theta):=r_{i}(s_{i})\vect e_{r}(\theta)+z_{i}(s_{i})\vect e_{z},
     \end{aligned}
     \right.
    \qquad
    \quad (s_{i}, \theta)\in I_{i}\times [0,2\pi],
    \label{eq:axisym_param}
\end{equation}
where $I_i$ is a bounded interval and $s_{i}$ is a parameter, not necessarily the arc-length of the generating curve.

\begin{figure}
\centering{}\includegraphics[width=0.7\textwidth]{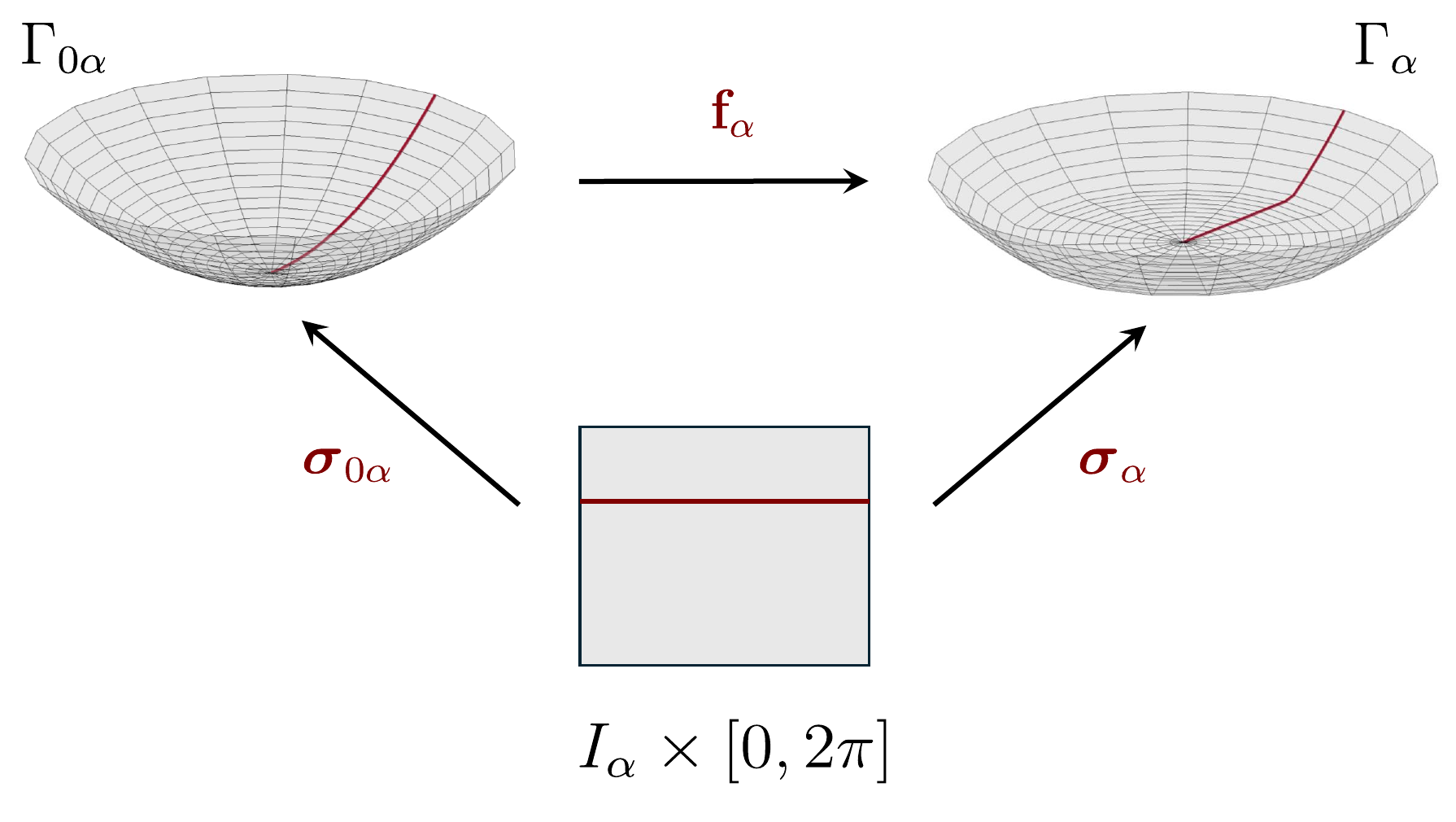}\caption{Commutative diagram explaining the parametrization adopted for the
reference and current curves, $\protect\vect{\sigma}_{0i}$ and $\protect\vect{\sigma}_{i}$,
respectively.\label{fig:Commutative-diagram}}
\end{figure}

Notice that the parameterization provided in Eq.\eqref{eq:axisym_param} also ensures the axisymmetry of the deformation map.
Indeed, as a consequence of such a parametrization, the commutative diagram
reported in Fig.\ref{fig:Commutative-diagram} holds. In particular, we can write the planar deformation gradient
as a function of $r_{0\alpha}$, $z_{0\alpha}$, $r_{\alpha}$ and $z_{\alpha}$ as
\begin{equation}
    \tens F_{\alpha}=\lambda_{\alpha s}\vect d_{\alpha s}\otimes\vect D_{\alpha s}+\lambda_{\alpha\theta}\vect d_{\alpha \theta}\otimes\vect D_{\alpha\theta},
\end{equation}
where the $\vect{D}$'s and the $\vect{d}$'s  are the unitary coordinate vectors tangent to the reference and current profiles, respectively. Moreover, the principal strains, denoted by $\lambda_{\alpha r}$ and $\lambda_{\alpha s}$, can be computed as
\[
\lambda_{\alpha\theta}(s_\alpha)=\frac{r_{\alpha}(s_{\alpha})}{r_{0\alpha}(s_{\alpha})},\qquad
\lambda_{\alpha s}(s_{\alpha})=\frac{\sqrt{r_{\alpha}'(s_{\alpha})^{2}+z_{\alpha}'(s_{\alpha})^{2}}}{\sqrt{r_{0\alpha}'(s_{\alpha})^{2}+z_{0\alpha}'(s_{\alpha})^{2}}}.
\]

Finally, let us introduce  $\vect{\sigma}_{0i},\vect{\sigma}_{i}: I_i \to \mathbb{R}^2$, where $\vect{\sigma}_{0i}=(r_{0i},z_{0i})$ and $\vect{\sigma}_i=(r_i,z_i)$. We define the space of admissible profiles $\mathscr{V}$ as
\begin{equation}
\begin{split}
    \mathscr{V}:=&\{\vect{\sigma}_i=(r_i,z_i),\,\,i=1,2,3\,|\, r_i = r_j  \text{ and } z_i=z_j \text{ at the triple contact line, }\\
    &r_1=r_2=0 \text{ at the symmetry axis and }r_3=R \text{ at the container's wall}  \}
\end{split}
\end{equation}

\subsection{Equilibrium equations for the reference profile}

In the undeformed reference configuration, only surface tension acts at the interface between the phases, as the newly formed membrane is considered to be in a relaxed state with no contribution from elastic stresses. 
The assumption of a stress-free membrane is motivated by the fact that adsorption of HFBII molecules occurs progressively during evaporation, allowing the interface to relax locally as the film forms. As a result, the reference configuration of the membrane can be approximated as stress-free, while subsequent deformation induces tensile or compressive stresses. Hence, among all admissible profiles the ones representing the equilibrium of the reference configurations must fulfill the following modified Young-Laplace equations
\begin{equation}
    \begin{array}{lll}
        \label{eq:reference_profile_eq}
        & \gamma_1 \kappa_{01}  =  - \rho_{\mathrm{d}}gz_{01} + p_0 ,
        & \text{on } \Gamma_{01}, \\
        & \gamma_2 \kappa_{02} =  (\rho_\d - \rho_\ell)g z_{02} -p_0 + q_0 ,
        & \text{on } \Gamma_{02},\\
        & \gamma_3 \kappa_{03} =  \rho_\ell g z_{03} - q_0, & \text{on } \Gamma_{03},
\end{array}
\end{equation}      
where $\rho_{\mathrm{d}}$ and $\rho_{\ell}$ are the mass densities of the droplet and the oil, respectively. The structure of the above equations is the following: the product of twice the reference mean-curvature $\kappa_{0i}$ of the $i-th$ profile with the surface tension $\gamma_i$ equals the pressure jump a the interface. The latter is given as the sum of two main contributions, the one given by gravitational forces involving the gravitational constant $g$ and the one given by the constants $p_0$ and $q_0$. They represent pressure jumps at the interfaces $\Gamma_{01}$ and $\Gamma_{03}$, respectively, and they arise as a byproduct of the incompressibility of the fluids. Indeed, as will be clear in the following, they are determined by requiring that the volumes of the droplet and of the oil are preserved.
The equations in Eq.\eqref{eq:reference_profile_eq} must be complemented with modified Neumann relations at the triple contact line and at container's wall, dictating the balance of forces at the borders of the interfaces. Specifically, at the triple contact line we get 
\begin{align}
    \label{eq:neuman_TP}
    \theta_1 + \theta_2 = \pi -\cos^{-1}\left( \frac{\gamma_1^2 + \gamma_2^2 -\gamma_3^2}{2 \gamma_1 \gamma_2} \right), \qquad 
    \theta_1 + \theta_3 = \cos^{-1}\left( \frac{\gamma_3^2 + \gamma_1^2 -\gamma_2^2}{2 \gamma_1 \gamma_3} \right),
\end{align}
with $\theta_i$ being the angles depicted in Fig.\ref{fig:sketch-droplet}. Instead, at the wall we have
\begin{equation}
     \gamma_3 \cos \theta_\mathrm{w} = \gamma_\mathrm{a} - \gamma_\ell,
     \label{eq:neuman_wall}
\end{equation}
where $\theta_\mathrm{w}$ is the contact angle with the wall, as seen in Fig.\ref{fig:sketch-droplet}, while $\gamma_\ell$ and $\gamma_\mathrm{a}$ are the surface tensions at the interfaces wall-oil and wall-air, respectively. 

\subsection{Current profiles}
We chose to describe the elastic film by means of a {membrane model} since the capsule is supposed to be very thin, with negligible bending rigidity.
Consequently, the equilibrium equations for the current profile are analogous to the ones reported in Eq.\eqref{eq:reference_profile_eq} except for the contribution of the elastic membrane that encapsulates the droplet. The equations for the {transverse equilibrium} read \citep{de1985wetting,walker2015shapes,wong2017non}
\begin{align}
\label{eq:current_profile_transverse_eq}
\begin{array}{lll}
    &\tens{\tau}_{1s}\tens{\kappa}_{1s} + \tens{\tau}_{1\theta}\tens{\kappa}_{1\theta}  = - \rho_{\mathrm{d}}gz_1 + p ,
    & \text{on } \Gamma_{1}, \\
    & \tens{\tau}_{2s}\tens{\kappa}_{2s} + \tens{\tau}_{2\theta}\tens{\kappa}_{2\theta} =   (\rho_\d - \rho_\ell)g z_2 -p +q, & \text{on } \Gamma_{2} ,\\
    & \gamma_3 \kappa_3 =  \rho_\ell g z_3 - q ,
    & \text{on } \Gamma_{3},
\end{array}
\end{align}
where $\tau_{\alpha s}$ and $\tau_{\alpha \theta}$ are the meridional and circumferential principal elastic stresses while $\kappa_{\alpha s}$ and $\kappa_{\alpha \theta}$ denote the principal curvatures of the profiles. The tensions $\tens{\tau}_{\alpha s}$ and $\tens{\tau}_{\alpha \theta}$ are found by solving additional equations for the {in-plane equilibrium} of the membrane, namely
\begin{equation}
    \frac{1}{r\sqrt{g_s}} \frac{\d}{\d s}(r \sqrt{g_s}\tau_s)
    -\frac{r'}{r}\tau_\theta
    - \frac{g_s'}{2g_s}\tau_s =0,
    \label{eq:in_plane_memebrane_eq}
\end{equation}
where $g_s:=(r')^2+(z')^2$ and the left-hand side of Eq.\eqref{eq:in_plane_memebrane_eq} is nothing but the superficial divergence of the Cauchy stress tensor specialized for a surface of revolution.
Once again, the aforementioned equations must be complemented with slightly different Neumann's like conditions at the contact lines. They read
\begin{align}
    \label{eq:cur_neuman_TP}
    \theta_{1} + \theta_2 = \pi -\cos^{-1}\left( \frac{\tau_{1s}^2 + \tau_{2s}^2 -\gamma_3^2}{2 \tau_{1s} \tau_{2s}} \right) \qquad 
    \theta_1 + \theta_3 = \cos^{-1}\left( \frac{\gamma_3^2 + \tau_{1s}^2 -\tau_{2s}^2}{2 \tau_{1s} \gamma_{3}} \right),
\end{align}
where in place of $\gamma_\alpha$ we find the meridional tensions $\tau_{\alpha s}$. Instead,  Eq.\eqref{eq:neuman_wall} remains unchanged.

\subsubsection{Constitutive assumptions}
We consider the following classical linear constitutive relation establishing the following relation for the principal Cauchy stresses
\begin{equation}
    \tau_{\alpha s}=\frac{E}{(1-\nu^{2})\lambda_{\alpha\theta}}\left(\epsilon_{\alpha s}+\nu \epsilon_{\alpha\theta}\right) \qquad
    \tau_{\alpha\theta}=\frac{E}{(1-\nu^{2})\lambda_{\alpha s}}\left(\epsilon_{\alpha\theta}+\nu \epsilon_{\alpha s}\right),
\label{eq:tau_tense}
\end{equation}
where $\epsilon_{\alpha s}=\lambda_{\alpha s} - 1$ and $\epsilon_{\alpha \theta}=\lambda_{\alpha \theta} - 1$.
However, the equations above cannot be considered valid for all the configurations. Indeed, due to the lack of bending rigidity, the film cannot sustain compressive stresses which are almost entirely relaxed through the formation of wrinkles. In the circumferentially wrinkled and crumpled regimes the membrane develops a non-axisymmetric out-of-plane microstructure; nevertheless, its coarse-grained mechanics can still be captured by an axisymmetric pseudo-surface obtained by averaging over the wrinkles, which is the surface described in our model, see~\citep{knoche2013elastometry}.

In our specific case, compressive regions arise as the droplet evaporates and its volume decreases, making it necessary to account for this phenomenon.
Specifically, under the simplifying assumption of fully developed wrinkles and completely relaxed compressive stresses, the axisymmetric case under consideration leads to exactly four mutually exclusive scenarios \citep{knoche2013elastometry}. In the first scenario, the membrane is taut in every direction and the relations in Eq.~\eqref{eq:tau_tense} hold. In the second one the membrane goes under compression only in the hoop direction and circumferential wrinkling occur. Here, we relax the hoop stress by setting $\tau_\theta=0$. This allows us to recover the relation $\lambda_\theta = \lambda_\theta(\lambda_s)$, which in turn enables us to express $\tau_s$ solely in terms of $\lambda_s$. The third scenario is analogous and dual to the second one as the membrane goes under compression in the meridional direction and remains taut in the other one. We refer to this case as to meridional wrinkling. In the fourth and final case both stresses are relaxed and set to zero. The membrane is said to be in a crumpling regime. In formulas, the stresses in each of the four scenarios are given by

\begin{equation}
\label{eq:stresses}
\begin{array}{lll}
    \tau_{\alpha s}=\frac{E\left(\epsilon_{\alpha s}+\nu \epsilon_{\alpha\theta}\right)}{(1-\nu^{2})\lambda_{\alpha\theta}} + \gamma_\alpha ,
    &\tau_{\alpha\theta}=\frac{E\left(\epsilon_{\alpha s}+\nu \epsilon_{\alpha\theta}\right)}{(1-\nu^{2})\lambda_{\alpha s}} + \gamma_\alpha,
    & \text{Taut membrane}, \\
    \tau_{\alpha s}=E \epsilon_{\alpha s} + \gamma_\alpha (1-\nu) ,
    & \tau_{\alpha \theta} = 0,
    & \text{Circumferential  wrinkling}, \\
     \tau_{\alpha s}=0 ,
    & \tau_{\alpha \theta} = E \epsilon_{\alpha \theta} + \gamma_\alpha (1-\nu) ,
    & \text{Meridional wrinkling},\\
    \tau_{\alpha s} = 0 ,
    & \tau_{\alpha \theta} = 0,
    & \text{Crumpling}.
\end{array}   
\end{equation}

\subsubsection{Nondimensionalization}
We take as characteristic length of the problem $\ell=V_{0\d}^{1/3}$  with $V_{0\d}$ the volume of the droplet in the reference state, and
non-dimensionalize the energy ${\cal E}$ by as $\gamma_{3}\ell^{2}$. The problem depends on the following non-dimensional parameters:
\begin{equation}
    \tilde{\gamma}_{\alpha}=\frac{\gamma_{\alpha}}{\gamma_{3}} \qquad
    \tilde{\gamma}_{\ell}=\frac{\gamma_{\ell}}{\gamma_{3}} \qquad
    \tilde{\gamma}_{\mathrm{a}}=\frac{\gamma_{\mathrm{a}}}{\gamma_{3}}
    \qquad D=\frac{\rho_{\mathrm{d}}}{\rho_{\mathrm{l}}}\qquad
    \mathrm{Bo}=\frac{\ell^{2}\rho_{\mathrm{l}}g}{\gamma_{3}}
\qquad \tilde{E}=\frac{E}{\gamma_3}
\end{equation}%
The parameter $D$ measures the density ratio between the droplet and the oil, $\tilde{E}$ measures the ratio between the membrane elasticity and the capillary forces while $\mathrm{Bo}$, also known as Bond number, the ratio between gravitational and capillary forces.
The non-dimensional equations then read
\begin{equation}
\begin{array}{ll}
\label{eq:non-dimesional_parameters}
    \tilde{\tens{\tau}}_{1s}\tens{\kappa}_{1s} + \tilde{\tens{\tau}}_{1\theta}\tens{\kappa}_{1\theta} + \mathrm{Bo} \, D \, z_1 - p = 0  
    & \text{on } \Gamma_{1} \\
    \tilde{\tens{\tau}}_{2s}\tens{\kappa}_{2s} + \tilde{\tens{\tau}}_{2\theta}\tens{\kappa}_{2\theta} - \mathrm{Bo} \, (D-1)\, z_2 + p - q =0 
    & \text{on } \Gamma_{2} \\
    \kappa_3  - \mathrm{Bo}\, z_3 - q =0
    & \text{on } \Gamma_{3}
\end{array}
\end{equation}
where the the expression on the $\tilde{\tau}$'s is identical to the one reported in Eq.\eqref{eq:stresses}, provided we replace $E$ and $\gamma$ with their non-dimensional counterparts in Eq. \eqref{eq:non-dimesional_parameters}.
In the following we will always
consider the non-dimensionalized problem and, to lighten the notation,
we will drop all the tildes on non-dimensional quantities, except
where explicitly mentioned.

\subsection{Variational formulation}

The numerical solution of the equilibrium equations for an axisymmetric droplet is typically addressed by means of a shooting method. However, this approach has so far been applied only to much simpler systems that involve very few shooting parameters \citep{riccobelli2023flattened,wong2017non}. Despite its conceptual simplicity, the method is numerically stiff, and struggles to solve the set of equations in \eqref{eq:reference_profile_eq} and \eqref{eq:current_profile_transverse_eq} due to the large number of shooting parameters involved. In this section, we therefore propose a variational formulation of the problem, based on the minimization of a free-energy functional, which leads to a more efficient and robust numerical method.

\subsubsection{Reference profile: variational formulation}
\label{refp}

The reference configuration ${\cal M}_0$ can be characterized as a stationary point of a free energy functional, $\mathcal{E}_0$, subject to the constraint that the volume of the droplet and the oil are kept fixed. The {non-dimensional} free energy consists of two contributions,
\begin{equation}
    {\cal E}_0 = {\cal S}_0 - {\cal G}_0 
\end{equation}
${\cal S}_0$ is a surface energy term that accounts for the interactions at the interface ${\cal M}_0$ and with the wall of the container. It reads
\begin{equation}
    {\cal S}_{0}=\sum_{\alpha}\gamma_{\alpha}|
    \Gamma_{0\alpha}| +
    |\Gamma_{03}| 
    + \gamma_{\ell}|\Gamma_{0\ell}|  + \gamma_{\mathrm{a}}|\Gamma_{0\mathrm{a}}| 
\end{equation}
where $|\cdot|$ denotes the measure of the surface. 

As for ${\cal G}_0$, it represents the total gravitational potential reading
\begin{equation}
\label{eq:ref_grav_potential}
    {\cal G}_{0}=- \mathrm{Bo} D \int_{\Omega_{0\d}} z-\mathrm{Bo} \int_{\Omega_{0\ell}}z
\end{equation}
where, $\Omega_{0\d}$ is the region enclosed by profiles 1 and 2, namely, the droplet, while
$\Omega_{0\ell}$ is the region occupied by the oil.

Hence, the variational formulation for the reference profile reads
\begin{equation}
    \min_{\vect{\sigma}_{0i}\in \mathscr{V}}{\cal E}_{0}  \text{ subject to } \text{Vol}(\Omega_{0\d})=V_{0\d} \text{ and } \text{Vol}(\Omega_{0\ell})=V_{0\ell}
\end{equation}
where $\mathscr{V}$ denotes the space of {admissible profiles}, $\mathrm{Vol}(\cdot)$ denotes the volume of a set and, $V_{0\d}$ and $V_{0\ell}$ are the prescribed values for the volume of the droplet and of the oil, respectively.

\subsubsection{Current profile - variational problem}\label{currp}

The equilibrium configuration of the encapsulated droplet minimizes the following total free
energy ${\cal E}$:
\[
{\cal E}={\cal S}-{\cal G}
\]
where ${\cal G}$ is the gravitational potential and its expression
is identical to that in Eq. (\ref{eq:ref_grav_potential}), namely,
\[
{\cal G}=- \mathrm{Bo} \, D \int_{\Omega_{\d}} z-\mathrm{Bo} \int_{\Omega_{\ell}}z
\]
On the other hand, the surface energy ${\cal S}$ is different as it accounts for the elastic contribution due to the presence of the encapsulating
membrane. 
Specifically, we define ${\cal S}$ as 
\begin{equation}
    {\cal S}=\sum_{\alpha=1}^{2} \int_{\Gamma_{0\alpha}}W_{\alpha}(\tens C_{\alpha})
    + \gamma_3|\Gamma_{3}|
    + \gamma_\ell|\Gamma_{\ell}|
    + \gamma_\mathrm{a}|\Gamma_{\mathrm{a}}|
\end{equation}
where, as usual, the term proportional to the measure of the surfaces accounts for the surface tension while the contribution given by the droplet membrane is expressed in terms of an elastic energy density $W_{\alpha}$.

\subsubsection{Quasi-convexification of the elastic functional}

The elastic energy corresponding to the constitutive relations in Eq.~\eqref{eq:stresses} has the following expression
\begin{equation}
    W_\alpha=\frac{E_\alpha}{2(1-\nu^{2})}\left(\epsilon_{\alpha s}^{2}+\epsilon_{\alpha \theta}^{2}+2\nu\epsilon_{\alpha s}\epsilon_{\alpha \theta}\right) + \gamma_\alpha J,
\end{equation}
where $J$ is the ratio between the corresponding area elements of the current and reference states. However,  such an energy cannot support compressive states and makes the equilibrium variational problem ill-posed.
In order to deal with compressive phenomena, we follow \cite{mosler2008novel} and exploit a relaxed elastic energy by which compressive stretches produce null planar stresses.
Such a relaxation corresponds to a quasi-convexification of ${\cal E}$, i.e.
\begin{equation}
    {\cal E}^*= {\cal S}^* - {\cal G},
\end{equation}
where
\[
{\cal S}^*=\sum_{\alpha=1}^{2} \int_{\Gamma_{0\alpha}}W_{\alpha}^*(\tens C_{\alpha})
    + \gamma_3|\Gamma_{3}|
    + \gamma_\ell|\Gamma_{\ell}|
    + \gamma_\mathrm{a}|\Gamma_{\mathrm{a}}|
\]
\begin{equation}
W_{\alpha}^{*}(\tens C_{\alpha})=\min_{\delta\tens C_{\alpha}}W_{\alpha}(\tens C_{\alpha}+\delta\tens C_{\alpha})\label{eq:relaxed_energy}
\end{equation}
Here $\delta\tens C_{\alpha}$ must be of the form, $\delta\tens C_{\alpha}=a^{2}\vect n\otimes\vect n+b^{2}\vect m\otimes\vect m$ with $\vect n$ and $\vect m$ being the wrinkle directions. In our axisymmetric problem $\vect n = \vect{D}_{\alpha s}$ and $\vect m = \vect{D}_{\alpha \theta}$.

So, the relaxed variational problem describing the equilibrium of the
current profile becomes
\begin{equation}
    \label{eq:relaxed_cur_var_prob}
    \min_{\sigma_i \in \mathscr{V}}{\cal E}^{*}\text{ subject to } \mathrm{Vol}(\Omega_{\d})=V_{\d} \text{ and } \mathrm{Vol}(\Omega_{\ell})=V_{\ell}
\end{equation}
where, as before, $\mathscr{V}$ denote the space of admissible profiles.

\subsection{Numerical procedure}

First, let us detail the parameterization space for the profiles described through $r_i=r_i(s_i)$ and $z_i=z_i(s_i)$ with $s_i\in I_i$. Specifically, we  set $I_{1}=I_{2}=I_{3}=I=[0,1]$, which allows us to rely on just one mesh of the interval $[0,1]$. Moreover, we parameterize the profiles so that, for every index $i$, we have that $(r_i(1),z_i(1))$ must coincide at the triple contact line. %
In particular, we impose as Dirichlet boundary conditions $r_{1}(0)=r_{2}(0)=0$ and $r_3(0)=R$.
Instead, the matching of the profiles at the triple contact line becomes expressed as
$r_i(1)=r_j(1)$ and $z_i(1)=z_j(1)$.
The conditions already introduced for the current profiles $r_i$ and $z_i$ are also applied to reference profiles 
$(r_{0i},z_{0i})$. %

We find the equilibrium profiles by numerically  minimizing the total free energy subject to the volume constraints (see Eq.\eqref{eq:relaxed_cur_var_prob}) that we enforce through Lagrange multipliers.
Moreover, additional Lagrange multipliers are used to impose the boundary conditions
$r_i(1)=r_j(1)$ and $z_i(1)=z_j(1)$ at the triple contact line. 

Let us introduce the variational problem associated with the {reference} configuration and express it as the stationarity of the following Lagrangian functional $\mathcal{L}_0$:
\begin{align}
{\cal L}_{0}(\vect{\sigma}_{0i},p_{0},q_0,\vect{\Lambda}_{0})= {\cal E}_{0}(\vect{\sigma}_{0i})-p_{0}\left(\mathrm{Vol}(\Omega_{0\mathrm{d}})-V_{\mathrm{0d}}\right) -
q_{0}\left(\mathrm{Vol}(\Omega_{0\ell})-V_{\mathrm{0\ell}}\right)
-{\cal B}(\vect{\sigma}_{0i})[\vect{\Lambda}_{0}]\label{eq:LAG-ref}
\end{align}
where $p_0$, $q_0$ and $\vect{\Lambda}_0=(\lambda_{01}, \lambda_{02},\lambda_{03})$ are Lagrange multipliers and $\cal B$ is the following linear form
\[
{\cal B}(M_0)[\vect{\Lambda}_0]=\lambda_{1}(r_{1}(1)-r_{2}(1))+\lambda_{2}(r_{1}(1)-r_{3}(1))+\lambda_{3}(z_{1}(1)-z_{2}(1))+\lambda_{4}(z_{3}(1)-z_{1}(1))
\]
enforcing, via the Lagrange multiplier $\vect{\Lambda}_0$,
the matching of the profiles at the triple contact line.

In order to reduce the sensitivity to the initial guess of the solution
we approach the local minimum of ${\cal E}_{0}$ by means of an $H^{1}$-gradient
flow, namely, find $\vect{\sigma}_{0}\in \mathscr{V}$ such that
\begin{equation}
\sum_{i=1}^3\langle \, \partial_{t}\vect{\sigma}_{0i}(t),\delta \vect{\sigma}_{0i} \, \rangle_{H^{1}(I)}+D_{\vect{\sigma}_{0i},p_{0},q_0,\Lambda_{0}}{\cal L}_{0}[\delta \vect{\sigma}_{0i},\delta p_{0},\delta\vect{\Lambda}_{0}]=0\,\,\,\,\,\forall\delta \vect{\sigma}_{0i} \in\delta \mathscr{V},\,\,\forall\delta p_{0},\delta\vect{\Lambda}_{0}\label{eq:GF-ref}
\end{equation}
where $\delta \mathscr{V}=\{\delta \vect{\sigma}_{0i} = (\delta r_{0i},\delta z_{0i})|\delta r_{3}(0)=0,\,\,\delta r_{1}(0)=r_{2}(0)=0\}$ is the space of admissible variations. In addition, for each $\vect{u}_i=(r_i,z_i)$ and $\vect{v}_i=(\rho_i , \zeta_i)$ the $H^1$-scalar product is defined
\begin{equation*}
    \langle \, \vect{u},\vect{v} \,\rangle_{H^1(I)}= \int_0^1 \left( r_i(s)\rho_i(s) + r_i'(s)\rho_i'(s) \right) \d s +
    \int_0^1 \left( z_i(s)\zeta_i(s) + z_i'(s)\zeta_i'(s) \right) \d s 
\end{equation*}
The ensuing evolution of $\vect{\sigma}_{0i}$ in the parameter $t$ converges to a local minimum of $\mathcal{E}_0$ and is solved by means of backward Euler
scheme relying on the Newton method. As for the spatial discretization,
we adopted linear $P_{1}$-finite elements implemented in the open source
library \texttt{FEniCS}.

Notice that, the stationarity of the Lagrangian in Eq. (\ref{eq:LAG-ref})
only determines the support of the planar curves parametrized by $\vect{\sigma}_{i}(s)$. Indeed, the definition of the reference free energy ${\cal E}_0$ only takes into account the {shape} of the profiles.
In practice, the specific parametrization $s_{i}\mapsto\vect{\sigma}_{0i}(s_{i})$
generated by the numerical solution of the equations is entirely determined
by the scalar product adopted in the definition of the gradient flow (in our case the $H^1$ scalar product).
Hence, the magnitude of $|\vect{\sigma}_{0\alpha}^{'}|$ is left essentially
uncontrolled and indeed can significantly vary across the domain ${\cal I}$
thus negatively affecting the numerical stiffness of the problem.
This is the reason why, in practice, instead of minimizing ${\cal E}_{0}$,
we minimize its augmented version ${\cal E}_{0}^{(\mathrm{aug})}$
defined as

\[
{\cal E}_{0}^{(\mathrm{aug})}(\vect{\sigma}_{0i})={\cal E
}_{0}(\vect{\sigma}_{0i})+\epsilon\sum_{i=1}^{3}(|\vect{\sigma}_{0i}^{'}|^{2}-1)
\]
where, the additional term proportional to $\epsilon$ penalizes deviations
from arc-length parametrization of the profiles thus improving the
stability of the numerical algorithm.

The current profile is determined in a completely analogous way, namely, find $\vect{\sigma}_{i}\in\mathscr{V}$ such that
\[
\sum_{i=1}^{3}\langle\,\partial_{t}\vect{\sigma}_{i},\delta\vect{\sigma}_{i}\,\rangle_{H^{1}({\cal I})}+D_{\vect{\sigma}_{i},p,q,\vect{\Lambda}}{\cal L}[\delta\vect{\sigma}_{i},\delta p,\delta\vect{\Lambda}]=0\,\,\,\,\,\forall\delta\vect{\sigma}_{i}\in\delta\vect V,\,\,\forall\delta p,\delta\vect{\Lambda}
\]
where 
\[
{\cal L}(\vect{\sigma}_{i},p,q,\vect{\Lambda})={\cal E}^{*}(\vect{\sigma}_{i})-p\left(\mathrm{Vol}(\Omega_{\mathrm{d}})-V_{\mathrm{d}}\right)
-q\left(\mathrm{Vol}(\Omega_{\ell})-V_{\ell}\right)
-{\cal B}(\vect{\sigma}_{i})[\vect{\Lambda}]
\]
Notice that, in such a case, the mapping $s_{\alpha}\mapsto\vect{\sigma}_{\alpha}(s_{\alpha})$
is uniquely determined without the need to augment the free energy
with any penalty term. Indeed, because of the presence of elastic energy, the shape of the final profiles is determined by an injective deformation map that establishes a one-to-one correspondence between points in the reference and current configurations.

\section{Results}

In the following, we summarize the results of the numerical simulations performed in order to investigate the evolution of the shape morphology over time during the different sets of evaporation experiments.

\begin{table}[b]
\begin{centering}
\begin{tabular}{ccc}
\toprule 
Dimensional quantity & Value & Reference\tabularnewline
\midrule
\midrule 
$\gamma_{1}$ & $0.055\,\,[\mathrm{mN}/\mathrm{m}]$ & \citep{riccobelli2023flattened}\tabularnewline
\midrule
$\gamma_{2}$ & $0.032\,\,[\mathrm{mN}/\mathrm{m}]$ & Estimated\tabularnewline
\midrule 
$\gamma_{3}$ & $0.022\,\,[\mathrm{mN}/\mathrm{m}]$ & \citep{SyensqoFomblinYLVAC256}\tabularnewline
\midrule 
$E$ & $400\,\,[\mathrm{mN}/\mathrm{m}]$ & \citep{riccobelli2023flattened}\tabularnewline
\midrule 
$\nu$ & $0.5 \text{ [\,-\,]}$ & \citep{riccobelli2023flattened}\tabularnewline
\midrule 
$\theta_{\mathrm{w}}$ & $-\pi/25 \text{ [rad]}$ & \text{Estimated}\tabularnewline
\bottomrule
\end{tabular}
\par\end{centering}
\caption{Physical parameters. We report the values of the physical parameters extracted by literature and estimated from  the experimental data.\label{tab:List-of-parameters}}
\end{table}

\subsection{Shape morphing prediction during evaporation against contact angle experiments}
\label{sec:shape-morphing-exp}

We begin by reproducing numerically the contact-angle experiment reported in Fig.~\ref{exp1}.
In this configuration the cuvette is filled up to the brim, so that the outer meniscus is small
and the observed shape is mainly governed by the balance between gravity and interfacial forces.
From the image acquired immediately after deposition we extract the axisymmetric drop contour and
use it to calibrate the {reference} equilibrium configuration by solving the variational problem
in Sec.~\ref{refp}. As suggested by the experimental time series (Fig.~\ref{exp1}), the profile remains
essentially unchanged during the early stage of the experiment, consistently with a quasi-static regime.
We therefore take this initial shape as a proxy for the stress-free (reference) configuration at the onset
of encapsulation.

\begin{figure}
    \centering
    \includegraphics[width=0.95\textwidth]{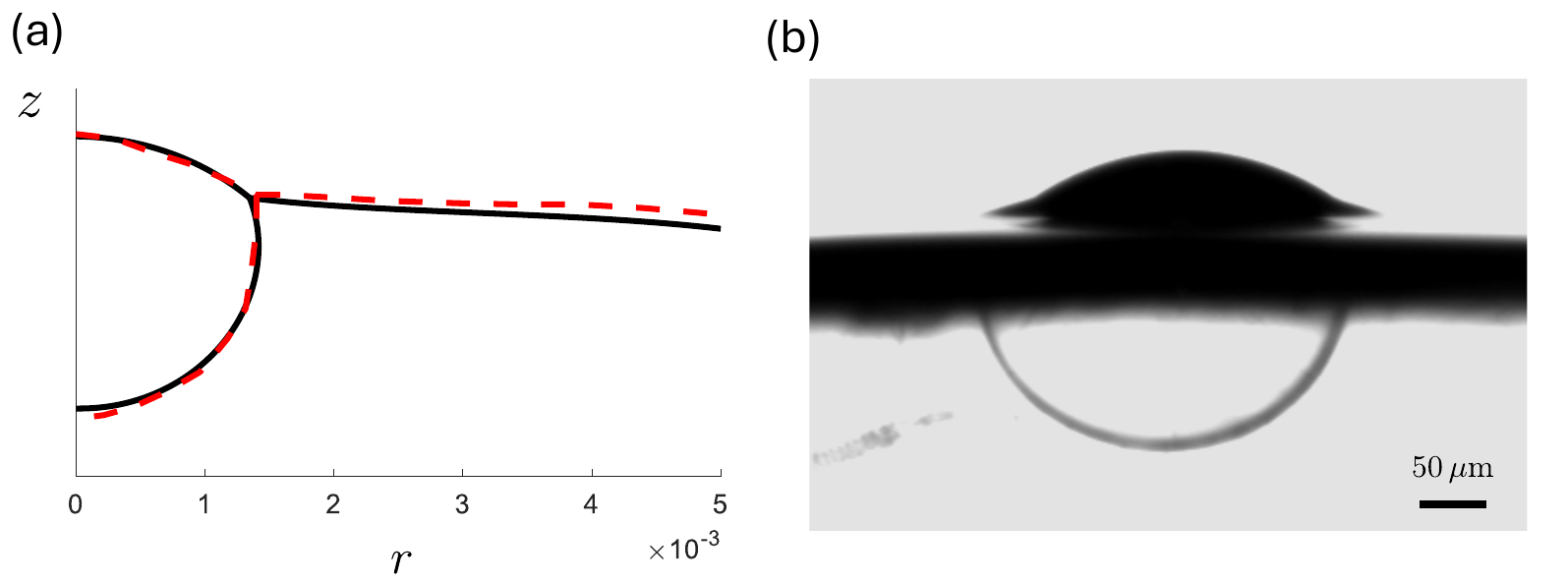}
    \caption{Reference profile used for parameter identification. (a) Comparison between experimental data
    (red dashed line) and theoretical prediction (black solid line). (b) Experimental image of the droplet
    before the onset of visible shape changes.}
    \label{fig:ref_profile_fit}
\end{figure}

In the fitting procedure we keep fixed the parameters taken from the literature
($\gamma_1$, $\gamma_3$, $E$, $\nu$) and estimate the oil--droplet surface tension $\gamma_2$
together with the effective contact angle at the container wall, $\theta_{\mathrm{w}}$,
by minimizing the mismatch between the experimental and numerical profiles.
The resulting agreement is shown in Fig.~\ref{fig:ref_profile_fit}, while the corresponding
dimensional parameters are summarized in Table~\ref{tab:List-of-parameters}.

\begin{figure}
    \centering
    \includegraphics[width=0.99\textwidth]{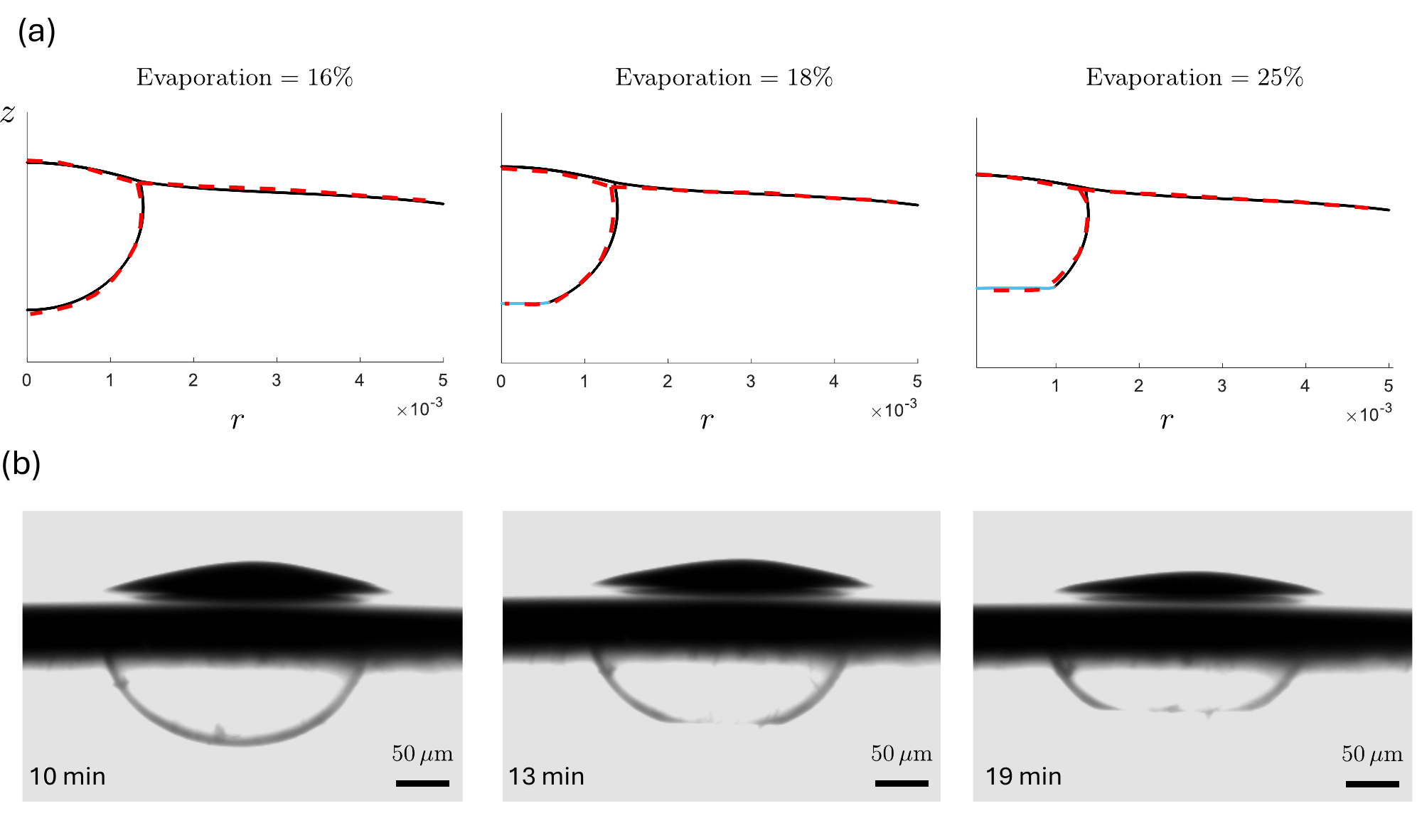}
    \caption{Evaporation-driven shape morphing. (a) Comparison between experimental profiles (red dashed)
    and numerical predictions (solid lines) for increasing evaporation. Black denotes taut membrane regions,
    while light blue denotes crumpled regions. (b) Corresponding experimental images (from left to right:
    16\%, 18\%, and 25\% volume loss).}
    \label{fig:evaporation_profiles}
\end{figure}

\begin{figure}
    \centering
    \includegraphics[width=0.9\textwidth]{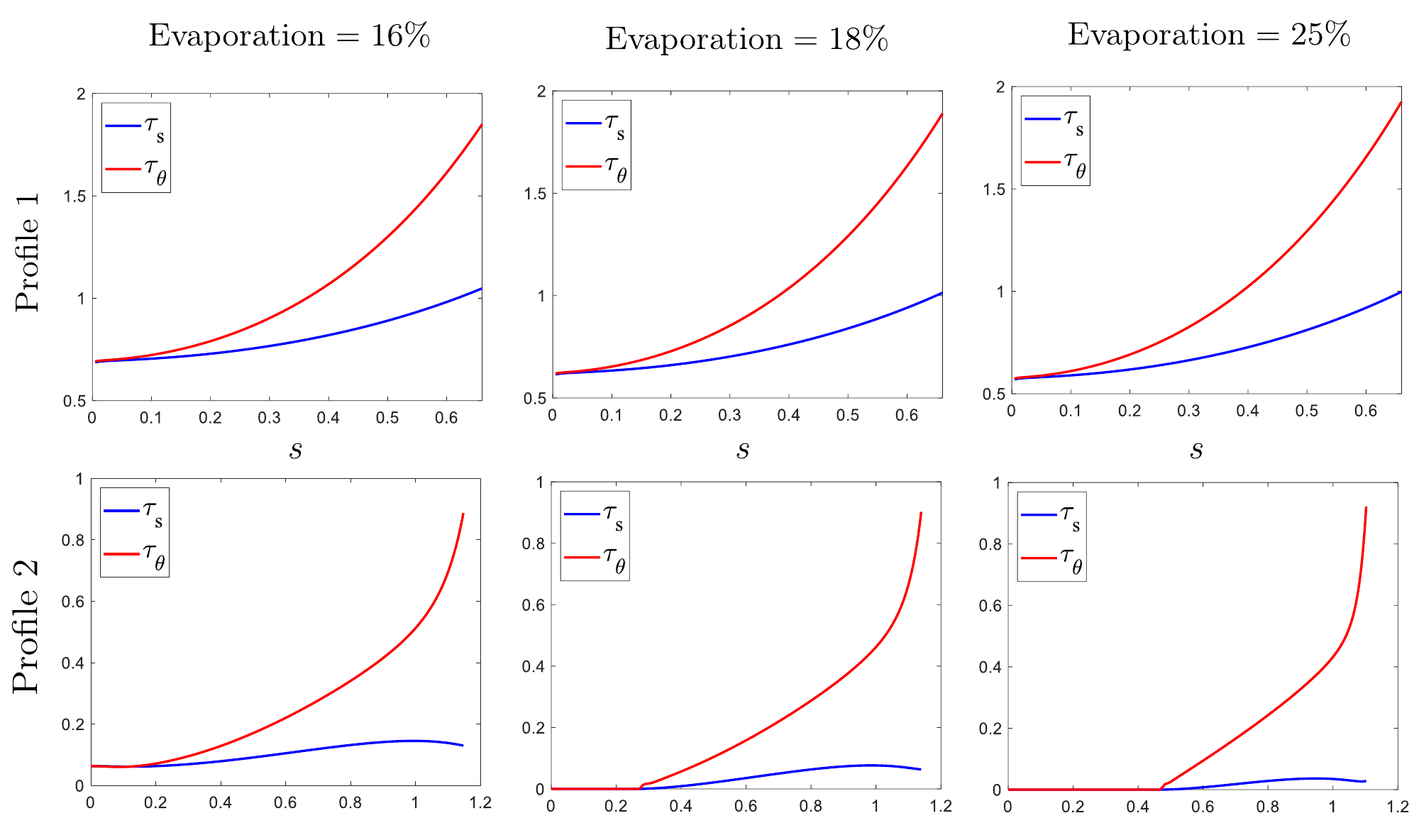}
    \caption{Meridional and hoop tensions, $\tau_s$ and $\tau_\theta$, as functions of the arc-length
    coordinate $s$ for $\Gamma_1$ and $\Gamma_2$ at increasing evaporation.}
    \label{fig:stress_profiles}
\end{figure}

After this initial calibration, we compute the subsequent quasi-static evolution driven by evaporation by
solving the {current} constrained minimization problem in Sec.~\ref{currp} at progressively smaller
droplet volumes. The reduced volume $\tilde V = V_{\mathrm d}/V_{0\mathrm d}$ is extracted from the images,
and for each value of $\tilde V$ we determine the equilibrium shape and the corresponding membrane state.
Figure~\ref{fig:evaporation_profiles} compares the predicted shapes with the experimental profiles for
three representative evaporation levels. The model captures quantitatively the progressive morphing of the
submerged interface and, in particular, the emergence and growth of a flattened region at the bottom of the
droplet--oil interface.

The underlying mechanical mechanism is clarified by the distributions of the principal tensions
$\tau_s$ and $\tau_\theta$ shown in Fig.~\ref{fig:stress_profiles}.
Along the droplet--air interface ($\Gamma_1$) the film remains in a tensile state throughout the process,
with both principal tensions positive. Conversely, along the droplet--oil interface ($\Gamma_2$) evaporation
induces a progressive loss of tension near the bottom apex: beyond a critical evaporation level the two
principal tensions simultaneously relax to zero, signalling the onset of a locally crumpled regime.
As evaporation proceeds, the crumpled patch expands upward, consistently with the experimentally observed
growth of the flattened area.

\subsection{Morphological phase diagrams}
\label{sec:phase_diagrams}

We next explore how the equilibrium morphology of an evaporating, self-encapsulated droplet depends on the control parameters governing the capillary--gravity balance and the interfacial tensions. We recall that the problem is governed by the following dimensionless quantities
\[
\tilde{\gamma}_{2}=\frac{\gamma_{2}}{\gamma_{3}},
\qquad
\tilde{V}=\frac{V_{\mathrm{d}}}{V_{0\mathrm{d}}},
\qquad
\mathrm{Bo}=\frac{\ell^{2}\rho_{\mathrm{l}}g}{\gamma_{3}},
\qquad
D=\frac{\rho_{\mathrm{d}}}{\rho_{\mathrm{l}}}.
\]
For visualization, we plot  the dimensionless reference profile ${\cal M}_0$ in blue, while the dimensionless current configuration is color-coded according to the relaxed membrane regime defined in Eq.~\eqref{eq:stresses}: taut (black), crumpled (light blue), circumferentially wrinkled (orange, $\tau_\theta=0$), and meridionally wrinkled (red, $\tau_s=0$).

\begin{figure}[!t]
\centering
\includegraphics[width=1\textwidth]{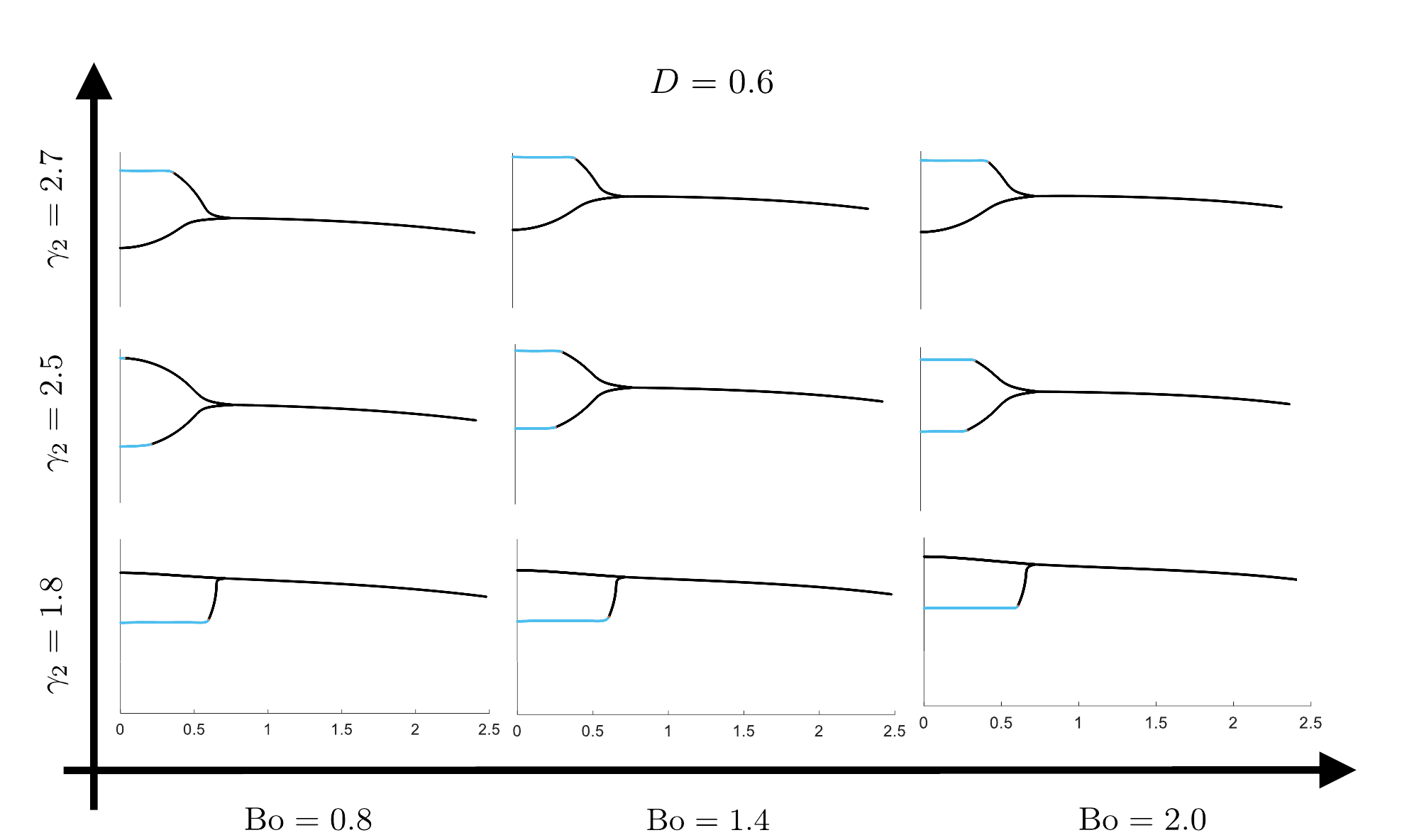}
\caption{Morphological diagram for $D=0.6$ at $\tilde{V}=0.4$ with $\tilde{\gamma}_1=2.5$, $\tilde{E}=18$ and $\theta_{\mathrm{w}}=-\pi/25$,
varying $\mathrm{Bo}$ (columns) and $\gamma_2$ (rows). Black: taut membrane. Light blue: isotropically crumpled regions ($\tau_s=\tau_\theta=0$),
which may appear on the exposed cap (upper branch), on the submerged cap (lower branch), or on both, depending on $(\mathrm{Bo},\gamma_2)$.}
\label{fig:phaseD06}
\end{figure}

\paragraph{Light droplet on a heavier oil ($D=0.6$): relocation of crumpling between the exposed and submerged caps}
We first consider a buoyant droplet ($D=0.6$) at fixed reduced volume $\tilde{V}=0.4$, with
$\tilde{\gamma}_1=2.5$, $\tilde{E}=18$, and $\theta_{\mathrm{w}}=-\pi/25$, while varying $\mathrm{Bo}$ and the
(non-dimensional) oil--droplet surface tension $\gamma_2$ (Fig.~\ref{fig:phaseD06}).
In this regime the instability manifests as isotropic crumpling (light-blue segments, i.e.\ $\tau_s=\tau_\theta=0$),
but the location of the crumpled patch depends on $(\mathrm{Bo},\,\gamma_2)$.

For large $\gamma_2$ (top row, $\gamma_2=2.7$) the crumpling is confined to the {exposed cap} (upper branch of the profile),
appearing as a small flattened portion near the symmetry axis, while the submerged interface remains fully taut.
This indicates that a strong oil--droplet tension penalizes deformation of the submerged cap, so that the excess area generated by
deflation is preferentially accommodated by relaxing the film on the exposed side.

Decreasing $\gamma_2$ progressively transfers the loss of tension to the {submerged cap} (lower branch).
At $\gamma_2=1.8$ (bottom row) crumpling is found exclusively below, where it produces an extended flat region at the droplet bottom. Thus, the instability patterns are confined to the submerged interface $\Gamma_2$: the droplet--air interface $\Gamma_1$ stays taut, in agreement with the stress analysis reported in Fig.~\ref{fig:stress_profiles}.
The intermediate case $\gamma_2=2.5$ (middle row) marks a crossover: at low gravity (small $\mathrm{Bo}$) crumpling is predominantly
on the exposed side, whereas increasing $\mathrm{Bo}$ promotes the emergence of an additional crumpled patch on the submerged side,
so that the relaxed region becomes {distributed on both caps}.
In general, increasing $\mathrm{Bo}$ favors tension loss at the submerged interface (by enhancing the sagging and hydrostatic pressure gradients),
while increasing $\gamma_2$ suppresses the bottom crumpling and shifts the relaxation towards the exposed cap.
In particular, within this $D=0.6$ scan we do not observe purely wrinkled states (single principal stress relaxation): the membrane either remains taut
or relaxes isotropically in localized crumpled patches.

\begin{figure}[!t]
\centering
\includegraphics[width=1\textwidth]{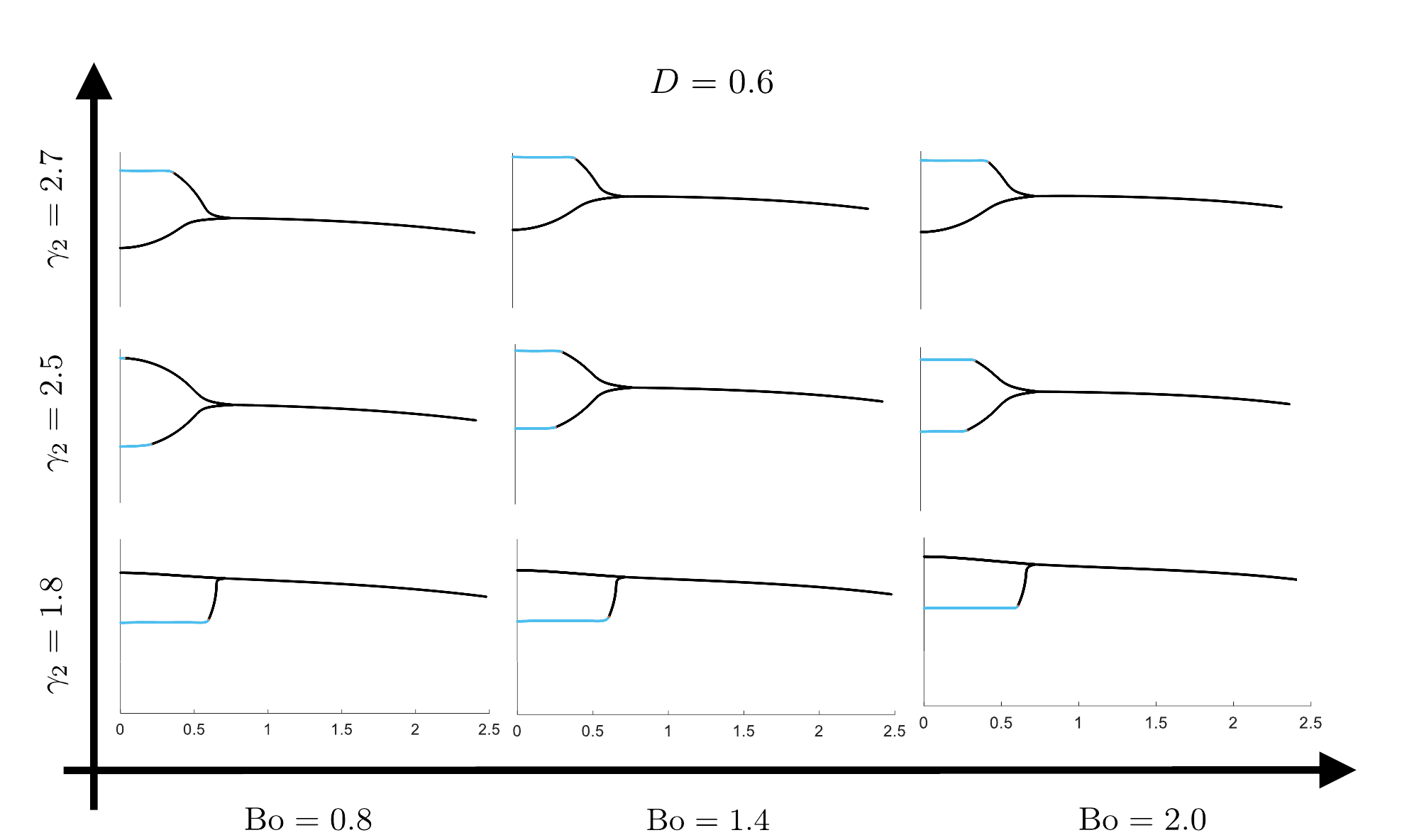}
\caption{Morphological diagram for $D=0.6$ at $\tilde{V}=0.4$ with $\tilde{\gamma}_1=2.5$, $\tilde{E}=18$ and $\theta_{\mathrm{w}}=-\pi/25$,
varying $\mathrm{Bo}$ (columns) and $\gamma_2$ (rows). Black: taut membrane. Light blue: isotropically crumpled regions ($\tau_s=\tau_\theta=0$),
which may appear on the exposed cap (upper branch), on the submerged cap (lower branch), or on both, depending on $(\mathrm{Bo},\gamma_2)$.}
\label{fig:phase_D06}
\end{figure}

\begin{figure}[!h]
\centering
\includegraphics[width=1\textwidth]{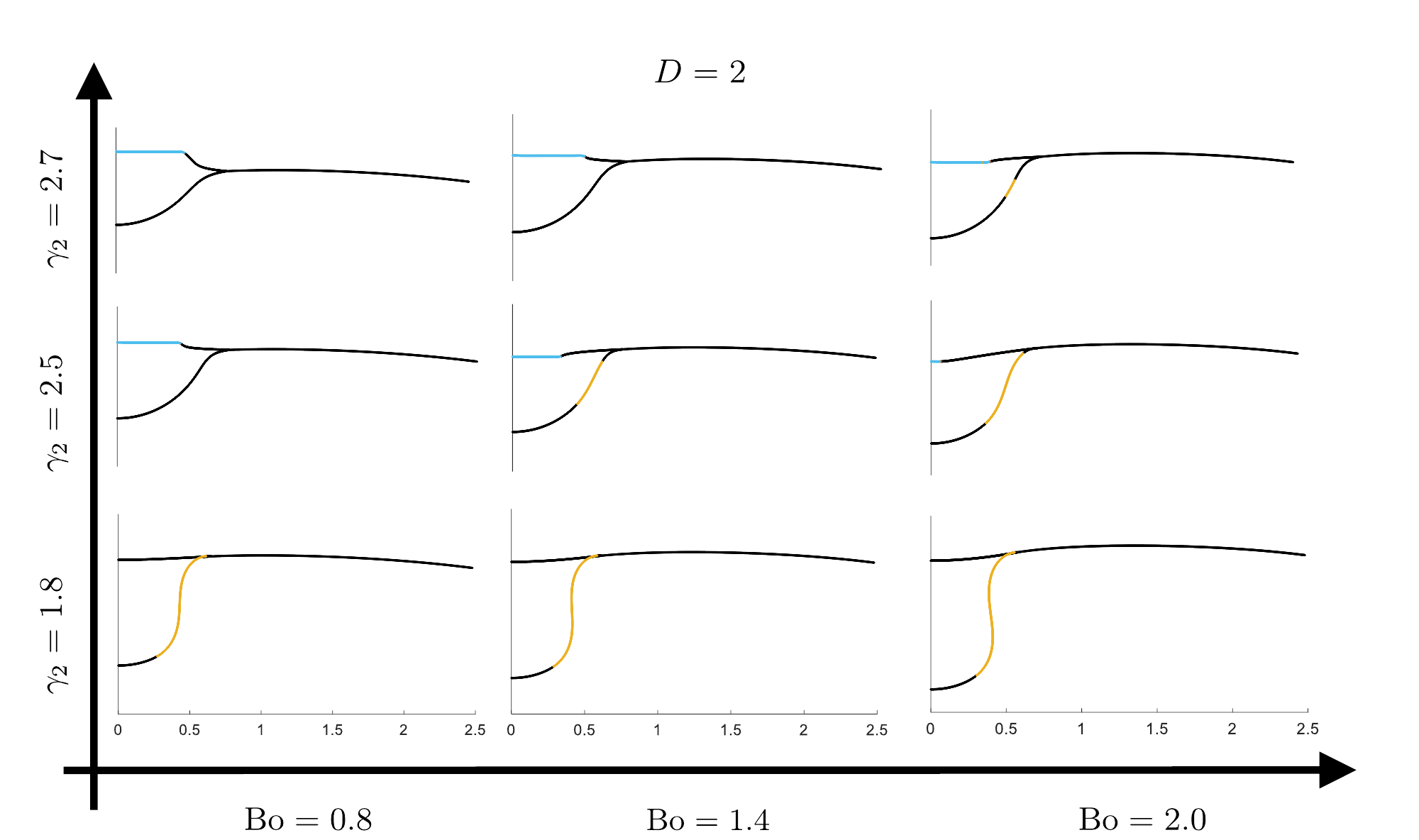}
\caption{Morphological diagram for $D=2$ at $\tilde{V}=0.4$ with $\tilde{\gamma}_1=2.5$, $\tilde{E}=18$ and $\theta_{\mathrm{w}}=-\pi/25$,
varying $\mathrm{Bo}$ (columns) and $\gamma_2$ (rows). Black: taut membrane.
Light blue: crumpled patches ($\tau_s=\tau_\theta=0$), mainly on the exposed cap at large $\gamma_2$.
Orange: circumferentially wrinkled regions ($\tau_\theta=0$), developing primarily along the submerged sidewall
and promoted by decreasing $\gamma_2$ and/or increasing $\mathrm{Bo}$.}
\label{fig:phase_D2}
\end{figure}

\paragraph{Heavy droplet in a lighter oil ($D=2$): crossover from exposed-cap crumpling to submerged-side wrinkling}
We now consider the opposite density contrast ($D=2$) at the same reduced volume $\tilde V=0.4$ and fixed
$\tilde{\gamma}_1=2.5$, $\tilde E=18$ and $\theta_{\mathrm w}=-\pi/25$, while varying $\mathrm{Bo}$ and $\gamma_2$
(Fig.~\ref{fig:phase_D2}).
In contrast with the buoyant case $D=0.6$, the relaxed states are no longer dominated by bottom crumpling:
the prevalent mechanism is {circumferential wrinkling} (orange, $\tau_\theta=0$), typically developing
along the {submerged lateral interface} $\Gamma_2$, i.e.\ on the steep ``sidewall'' portion of the lower branch.

The diagram reveals a clear two-step scenario controlled by $(\mathrm{Bo},\gamma_2)$.
For large $\gamma_2$ (top row, $\gamma_2=2.7$) the profile is mostly taut, except for a small {crumpled}
patch on the {exposed cap} (light blue near the symmetry axis). This indicates that a strong oil--droplet
tension stiffens the submerged interface, so that the residual compressive mismatch generated by deflation is
first accommodated on the exposed side. At sufficiently large gravity (right column, $\mathrm{Bo}=2.0$),
a short {wrinkled} segment also appears near the neck region, signalling the onset of hoop compression
when sagging becomes strong enough.

Decreasing $\gamma_2$ activates and enlarges a {wrinkled band} on $\Gamma_2$.
At $\gamma_2=2.5$ (middle row) the system undergoes a crossover with $\mathrm{Bo}$:
for small $\mathrm{Bo}$ the relaxation is still mainly on the exposed cap (crumpling), whereas increasing
$\mathrm{Bo}$ triggers a pronounced circumferentially wrinkled region on the submerged sidewall, which grows
upwards along $\Gamma_2$.
At low $\gamma_2$ (bottom row, $\gamma_2=1.8$) wrinkling becomes the dominant outcome for all $\mathrm{Bo}$:
the orange band occupies most of the submerged sidewall, while the exposed cap remains taut and the crumpled
patch disappears.
This confirms that, for heavy droplets, deflation preferentially generates {hoop compression} on the
highly curved immersed interface, leading to $\tau_\theta\to 0$ (wrinkling) rather than to an isotropic
stress collapse ($\tau_s=\tau_\theta=0$).

\begin{figure}[!t]
\centering
\includegraphics[width=1\textwidth]{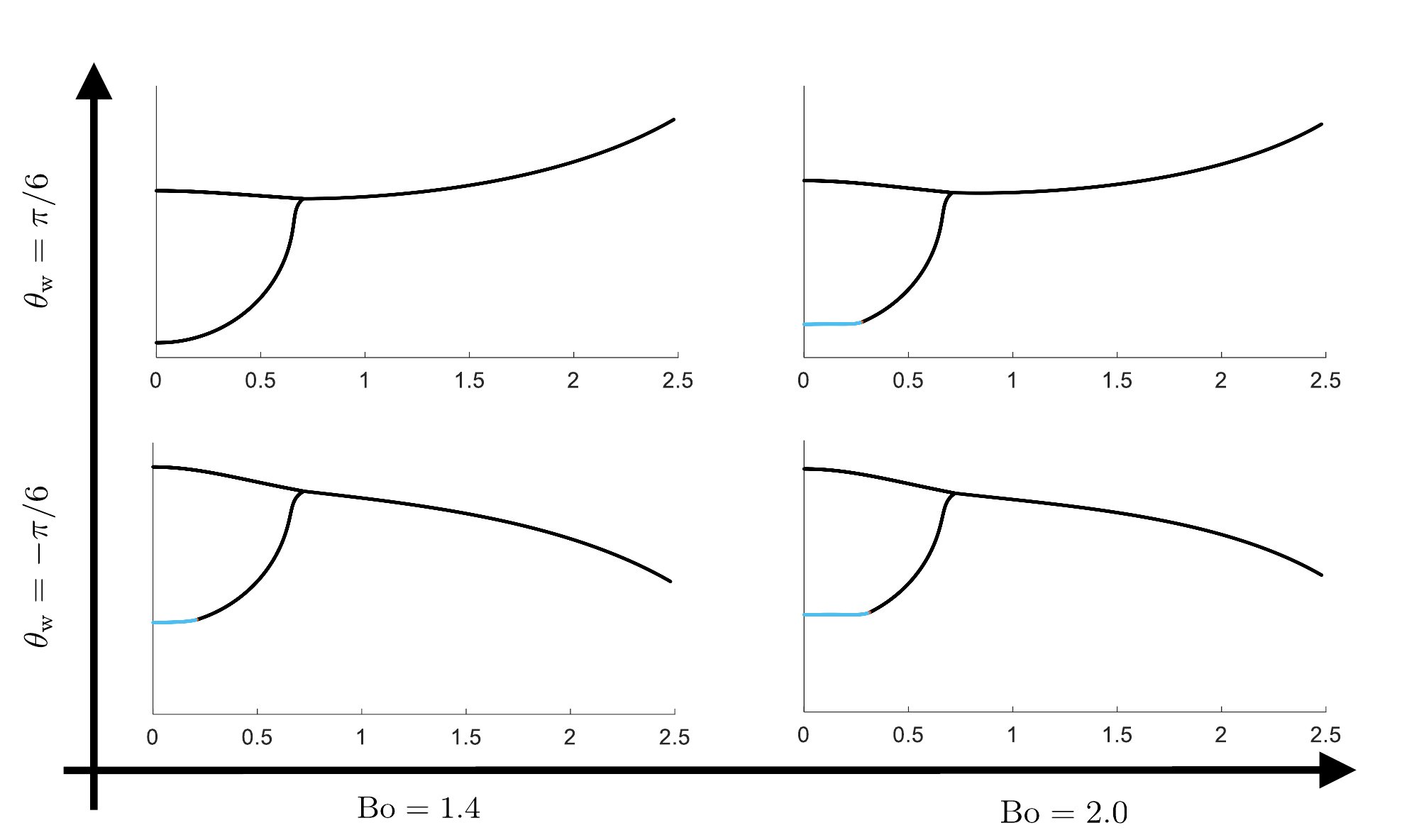}
\caption{Role of the wall meniscus controlled by the contact angle $\theta_{\mathrm w}$.
Profiles obtained by varying $\mathrm{Bo}$ and $\theta_{\mathrm w}$ at fixed
$\tilde{\gamma}_1=2.5$, $\tilde{\gamma}_2=1.8$, $\tilde{V}=0.69$, $D=0.6$, and $\tilde{E}=18$.
Increasing the meniscus (through $\theta_{\mathrm w}$) progressively suppresses the crumpled region on $\Gamma_2$,
leading to a fully taut submerged interface in the last configuration, in qualitative agreement with Fig.~\ref{exp2}.}
\label{fig:meniscus_theta}
\end{figure}

\paragraph{Meniscus control at the wall: increasing the meniscus suppresses crumpling}
Finally, we investigate the role of the outer meniscus at the container wall by varying the wall contact angle
$\theta_{\mathrm w}$ (Fig.~\ref{fig:meniscus_theta}) at fixed $\tilde{\gamma}_1=2.5$, $\tilde{\gamma}_2=1.8$,
$\tilde V=0.69$, $D=0.6$, and $\tilde E=18$.
Changing $\theta_{\mathrm w}$ alters the curvature and elevation of the oil--air interface near the wall and therefore
the position of the contact line and the capillary pressure level in the oil domain.
This has a direct mechanical consequence on the submerged interface $\Gamma_2$: as the meniscus becomes more pronounced,
the capsule is effectively pulled upward near the wall and the immersed cap is put under stronger in-plane stretch.
As a result, the crumpled patch on $\Gamma_2$ progressively shrinks and can be completely removed (top-left panel),
where the membrane remains fully taut. Notably, this last configuration reproduces {qualitatively} the experimental situation shown in Fig.~2:
when the droplet interacts more strongly with the wall meniscus, the bottom flattening/crumpling observed in the
axisymmetric configuration is suppressed. In this sense, the wall meniscus acts as an additional control parameter
that can switch off crumpling by restoring tensile stresses on $\Gamma_2$.

\section{Discussion and conclusions}
\label{sec:discussion_conclusions}

In this work we addressed the mechanics of an evaporating, self-encapsulated floating droplet, a deceptively simple system in which a thin elastic film, multiple interfaces, gravity and confinement interact to generate non-trivial shape transitions. Our aim was to provide a predictive framework capable of connecting experimentally observable morphologies to measurable physical controls (volume loss, interfacial tensions, density contrast and capillary length scales), while retaining a transparent mechanical interpretation of the stress pathways underlying the morphological transitions.

Methodologically, the core originality of our approach is to treat the encapsulating film as an axisymmetric elastic membrane whose equilibrium is obtained from a fully coupled variational formulation, where capillarity, hydrostatics and membrane elasticity are minimized simultaneously under geometric and volume constraints. Crucially, we embed in this energetic framework a tension-field type relaxation that selects, pointwise along the interfaces, the physically admissible stress state: taut regions coexist with relaxed phases in which one or both principal membrane tensions collapse, giving rise to wrinkled (uniaxial relaxation) or crumpled (biaxial relaxation) domains. This provides a mechanically grounded approach allowing to infer where and how the film relieves compression and a morphological transition occurs. The results of finite element simulations allow direct, quantitative comparison with evaporation experiments and produces phase-diagram predictions in a parameter space that is experimentally accessible.

From a physics perspective, our results reveal a clear separation of roles between gravity and elastocapillarity in governing the morphological diagram. Gravity, quantified by the Bond number, primarily controls the extent and localization of tension loss by amplifying sagging and hydrostatic pressure gradients, thereby promoting relaxed regions on the submerged interface and increasing their spatial support. Elastocapillarity, encoded here through the ratios of interfacial tensions and the membrane stiffness, determines which interface pays the energetic cost of accommodating the mismatch generated by deflation: increasing the oil--droplet surface tension penalizes deformation of the immersed cap and can shift relaxation towards the exposed cap, whereas decreasing it makes the submerged interface energetically favorable and initiates bottom flattening and stress collapse. Density contrast further selects the relaxation pathway: buoyant droplets preferentially exhibit isotropic stress collapse (crumpling) localized at caps, while heavier droplets tend to develop hoop compression along the immersed sidewall and relax it through circumferential wrinkling. Finally, we showed that the outer meniscus at the wall is not a secondary geometric detail but an active control parameter: by modifying the capillary pressure level and the position of the triple line, a pronounced meniscus can restore tension on the submerged interface and suppress crumpling, reproducing qualitatively the wall-affected experimental morphologies.

The present model has limitations that suggest several research directions. First, we restrict to axisymmetry and to a membrane description; this excludes non-axisymmetric pattern selection (wavelength and amplitude of wrinkles) and neglects bending stiffness, which may become relevant near highly curved regions and sharp folds. Second, evaporation is treated quasi-statically through a prescribed volume decrease; coupling the mechanics to transport (evaporation kinetics, solute effects, Marangoni stresses) would enable fully predictive dynamics and time-to-instability estimates. Third, our representation of the relaxed phases is energetic and coarse-grained; incorporating a post-buckling model that resolves wrinkle microstructure, or extending the theory to shells with finite bending and possible plasticity/viscoelasticity, would broaden applicability to real encapsulants and biofilms. On the experimental side, systematic measurements of interfacial tensions during evaporation and controlled manipulation of the meniscus would further constrain the parameter identification and sharpen quantitative validation.

Despite these limitations, the framework introduced here is readily extensible and offers a versatile design tool for engineering and life-science applications where soft interfaces and confinement dominate morphology. In microfluidics and soft manufacturing, the ability to predict and control transitions between taut, wrinkled and crumpled states provides a route to tunable textures, adaptive optics and encapsulation strategies driven by controlled volume changes. In biological contexts, similar elasto--capillary instabilities arise when tissues or extracellular matrices interact with fluid interfaces and pressure gradients; one prospective direction is to couple our mechanical formulation to growth and active stresses, enabling minimal yet predictive models for morphological remodeling in living systems.

\section*{Acknowledgements}
{PC, DA and DR have been partially supported by MUR, PRIN Research Projects 2020F3NCPX and grant Dipartimento di Eccellenza 2023-2027. PC gratefully acknowledges the support from the Istituto Nazionale di Alta Matematica (INdAM) and the Gruppo Nazionale per la Fisica Matematica (GNFM). The research of  Pasquale Ciarletta is part of the activities of ”Dipartimento di Eccellenza 2023–2027”
(MUR, Italy), Dipartimento di Matematica, Politecnico di Milano. The VTT Technical Research Centre of Finland and Syensqo are acknowledged for kindly providing samples of HFBII and Fomblin, respectively.}

\bibliography{references}

\begin{thebibliography}{29}
\providecommand{\natexlab}[1]{#1}
\providecommand{\url}[1]{\texttt{#1}}
\expandafter\ifx\csname urlstyle\endcsname\relax
  \providecommand{\doi}[1]{doi: #1}\else
  \providecommand{\doi}{doi: \begingroup \urlstyle{rm}\Url}\fi

\bibitem[Abkarian et~al.(2013)Abkarian, Proti{\`e}re, Aristoff, and
  Stone]{abkarian2013gravity}
M.~Abkarian, S.~Proti{\`e}re, J.~M. Aristoff, and H.~A. Stone.
\newblock Gravity-induced encapsulation of liquids by destabilization of
  granular rafts.
\newblock \emph{Nature communications}, 4\penalty0 (1):\penalty0 1895, 2013.

\bibitem[Bala~Subramaniam et~al.(2005)Bala~Subramaniam, Abkarian, Mahadevan,
  and Stone]{bala2005non}
A.~Bala~Subramaniam, M.~Abkarian, L.~Mahadevan, and H.~A. Stone.
\newblock Non-spherical bubbles.
\newblock \emph{Nature}, 438\penalty0 (7070):\penalty0 930--930, 2005.

\bibitem[Basu et~al.(2016)Basu, Bansal, and Miglani]{basu2016towards}
S.~Basu, L.~Bansal, and A.~Miglani.
\newblock Towards universal buckling dynamics in nanocolloidal sessile
  droplets: the effect of hydrophilic to superhydrophobic substrates and
  evaporation modes.
\newblock \emph{Soft matter}, 12\penalty0 (22):\penalty0 4896--4902, 2016.

\bibitem[Burton et~al.(2010)Burton, Huisman, Alison, Rogerson, and
  Taborek]{burton2010experimental}
J.~C. Burton, F.~M. Huisman, P.~Alison, D.~Rogerson, and P.~Taborek.
\newblock Experimental and numerical investigation of the equilibrium geometry
  of liquid lenses.
\newblock \emph{Langmuir}, 26\penalty0 (19):\penalty0 15316--15324, 2010.

\bibitem[De~Gennes(1985)]{de1985wetting}
P.-G. De~Gennes.
\newblock Wetting: statics and dynamics.
\newblock \emph{Reviews of modern physics}, 57\penalty0 (3):\penalty0 827,
  1985.

\bibitem[Deegan et~al.(1997)Deegan, Bakajin, Dupont, Huber, Nagel, and
  Witten]{deegan1997capillary}
R.~D. Deegan, O.~Bakajin, T.~F. Dupont, G.~Huber, S.~R. Nagel, and T.~A.
  Witten.
\newblock Capillary flow as the cause of ring stains from dried liquid drops.
\newblock \emph{Nature}, 389\penalty0 (6653):\penalty0 827--829, 1997.

\bibitem[Desai and Schwendeman(2013)]{desai2013active}
K.-G.~H. Desai and S.~P. Schwendeman.
\newblock Active self-healing encapsulation of vaccine antigens in plga
  microspheres.
\newblock \emph{Journal of Controlled Release}, 165\penalty0 (1):\penalty0
  62--74, 2013.

\bibitem[Fu et~al.(2014)Fu, Liu, Adri{\`a}, Shao, Cai, and
  Chipot]{fu2014material}
H.~Fu, Y.~Liu, F.~Adri{\`a}, X.~Shao, W.~Cai, and C.~Chipot.
\newblock From material science to avant-garde cuisine. the art of shaping
  liquids into spheres.
\newblock \emph{The Journal of Physical Chemistry B}, 118\penalty0
  (40):\penalty0 11747--11756, 2014.

\bibitem[Given~Jr(2009)]{given2009encapsulation}
P.~S. Given~Jr.
\newblock Encapsulation of flavors in emulsions for beverages.
\newblock \emph{Current Opinion in Colloid \& Interface Science}, 14\penalty0
  (1):\penalty0 43--47, 2009.

\bibitem[Knoche et~al.(2013)Knoche, Vella, Aumaitre, Degen, Rehage, Cicuta, and
  Kierfeld]{knoche2013elastometry}
S.~Knoche, D.~Vella, E.~Aumaitre, P.~Degen, H.~Rehage, P.~Cicuta, and
  J.~Kierfeld.
\newblock Elastometry of deflated capsules: Elastic moduli from shape and
  wrinkle analysis.
\newblock \emph{Langmuir}, 29\penalty0 (40):\penalty0 12463--12471, 2013.

\bibitem[Langmuir(1933)]{langmuir1933oil}
I.~Langmuir.
\newblock Oil lenses on water and the nature of monomolecular expanded films.
\newblock \emph{The Journal of Chemical Physics}, 1\penalty0 (11):\penalty0
  756--776, 1933.

\bibitem[Lathia et~al.(2023)Lathia, Nagpal, Modak, Mishra, Sharma, Reddy,
  Nukala, Bhat, and Sen]{lathia2023tunable}
R.~Lathia, S.~Nagpal, C.~D. Modak, S.~Mishra, D.~Sharma, B.~S. Reddy,
  P.~Nukala, R.~Bhat, and P.~Sen.
\newblock Tunable encapsulation of sessile droplets with solid and liquid
  shells.
\newblock \emph{Nature Communications}, 14\penalty0 (1):\penalty0 6445, 2023.

\bibitem[Legland et~al.(2016)Legland, Arganda-Carreras, and
  Andrey]{legland2016morpholibj}
D.~Legland, I.~Arganda-Carreras, and P.~Andrey.
\newblock Morpholibj: integrated library and plugins for mathematical
  morphology with imagej.
\newblock \emph{Bioinformatics}, 32\penalty0 (22):\penalty0 3532--3534, 2016.

\bibitem[Li et~al.(2024)Li, Zhang, and Wang]{li2024evaporative}
W.~Li, C.~Zhang, and Y.~Wang.
\newblock Evaporative self-assembly in colloidal droplets: Emergence of ordered
  structures from complex fluids.
\newblock \emph{Advances in Colloid and Interface Science}, 333:\penalty0
  103286, 2024.

\bibitem[Mazzara et~al.(2019)Mazzara, Ochyl, Hong, Moon, Prausnitz, and
  Schwendeman]{mazzara2019self}
J.~M. Mazzara, L.~J. Ochyl, J.~K. Hong, J.~J. Moon, M.~R. Prausnitz, and S.~P.
  Schwendeman.
\newblock Self-healing encapsulation and controlled release of vaccine antigens
  from plga microparticles delivered by microneedle patches.
\newblock \emph{Bioengineering \& translational medicine}, 4\penalty0
  (1):\penalty0 116--128, 2019.

\bibitem[Mosler(2008)]{mosler2008novel}
J.~Mosler.
\newblock A novel variational algorithmic formulation for wrinkling at finite
  strains based on energy minimization: application to mesh adaption.
\newblock \emph{Computer Methods in Applied Mechanics and Engineering},
  197\penalty0 (9-12):\penalty0 1131--1146, 2008.

\bibitem[Mou et~al.(2020)Mou, Deng, Hu, Wang, Deng, Xiao, and
  Zhan]{mou2020controllable}
C.-L. Mou, Q.-Z. Deng, J.-X. Hu, L.-Y. Wang, H.-B. Deng, G.~Xiao, and Y.~Zhan.
\newblock Controllable preparation of monodisperse alginate microcapsules with
  oil cores.
\newblock \emph{Journal of Colloid and Interface Science}, 569:\penalty0
  307--319, 2020.

\bibitem[Pauchard and Allain(2003)]{pauchard2003mechanical}
L.~Pauchard and C.~Allain.
\newblock Mechanical instability induced by complex liquid desiccation.
\newblock \emph{Comptes rendus. Physique}, 4\penalty0 (2):\penalty0 231--239,
  2003.

\bibitem[Prasath et~al.(2021)Prasath, Marthelot, Menon, and
  Govindarajan]{prasath2021wetting}
S.~G. Prasath, J.~Marthelot, N.~Menon, and R.~Govindarajan.
\newblock Wetting and wrapping of a floating droplet by a thin elastic
  filament.
\newblock \emph{Soft Matter}, 17\penalty0 (6):\penalty0 1497--1504, 2021.

\bibitem[Py et~al.(2007)Py, Reverdy, Doppler, Bico, Roman, and
  Baroud]{py2007capillary}
C.~Py, P.~Reverdy, L.~Doppler, J.~Bico, B.~Roman, and C.~N. Baroud.
\newblock Capillary origami: spontaneous wrapping of a droplet with an elastic
  sheet.
\newblock \emph{Physical Review Letters}, 98\penalty0 (15):\penalty0 156103,
  2007.

\bibitem[Reinhold et~al.(2012)Reinhold, Desai, Zhang, Olsen, and
  Schwendeman]{reinhold2012self}
S.~E. Reinhold, K.-G.~H. Desai, L.~Zhang, K.~F. Olsen, and S.~P. Schwendeman.
\newblock Self-healing microencapsulation of biomacromolecules without organic
  solvents.
\newblock \emph{Angewandte Chemie}, 124\penalty0 (43):\penalty0 10958--10961,
  2012.

\bibitem[Riccobelli et~al.(2023)Riccobelli, Al-Terke, Laaksonen, Metrangolo,
  Paananen, Ras, Ciarletta, and Vella]{riccobelli2023flattened}
D.~Riccobelli, H.~H. Al-Terke, P.~Laaksonen, P.~Metrangolo, A.~Paananen, R.~H.
  Ras, P.~Ciarletta, and D.~Vella.
\newblock Flattened and wrinkled encapsulated droplets: Shape morphing induced
  by gravity and evaporation.
\newblock \emph{Physical Review Letters}, 130\penalty0 (21):\penalty0 218202,
  2023.

\bibitem[Syensqo(2025)]{SyensqoFomblinYLVAC256}
Syensqo.
\newblock Fomblin y lvac 25/6, 2025.
\newblock URL \url{https://www.syensqo.com/en/product/fomblin-y-lvac-256}.

\bibitem[Tsapis et~al.(2005)Tsapis, Dufresne, Sinha, Riera, Hutchinson,
  Mahadevan, and Weitz]{tsapis2005onset}
N.~Tsapis, E.~R. Dufresne, S.~S. Sinha, C.~S. Riera, J.~W. Hutchinson,
  L.~Mahadevan, and D.~A. Weitz.
\newblock Onset of buckling in drying droplets of colloidal suspensions.
\newblock \emph{Physical review letters}, 94\penalty0 (1):\penalty0 018302,
  2005.

\bibitem[Walker(2015)]{walker2015shapes}
S.~W. Walker.
\newblock \emph{The shapes of things: a practical guide to differential
  geometry and the shape derivative}.
\newblock SIAM, 2015.

\bibitem[Wong et~al.(2017)Wong, Adda-Bedia, and Vella]{wong2017non}
C.~Y. Wong, M.~Adda-Bedia, and D.~Vella.
\newblock Non-wetting drops at liquid interfaces: from liquid marbles to
  leidenfrost drops.
\newblock \emph{Soft Matter}, 13\penalty0 (31):\penalty0 5250--5260, 2017.

\bibitem[Wulsten et~al.(2009)Wulsten, Kiekens, van Dycke, Voorspoels, and
  Lee]{wulsten2009levitated}
E.~Wulsten, F.~Kiekens, F.~van Dycke, J.~Voorspoels, and G.~Lee.
\newblock Levitated single-droplet drying: Case study with itraconazole dried
  in binary organic solvent mixtures.
\newblock \emph{International journal of pharmaceutics}, 378\penalty0
  (1-2):\penalty0 116--121, 2009.

\bibitem[Yamasaki and Haruyama(2016)]{yamasaki2016formation}
R.~Yamasaki and T.~Haruyama.
\newblock Formation mechanism of flattened top hfbi domical droplets.
\newblock \emph{The Journal of Physical Chemistry B}, 120\penalty0
  (15):\penalty0 3699--3704, 2016.

\bibitem[Yamasaki et~al.(2016)Yamasaki, Takatsuji, Asakawa, Fukuma, and
  Haruyama]{yamasaki2016flattened}
R.~Yamasaki, Y.~Takatsuji, H.~Asakawa, T.~Fukuma, and T.~Haruyama.
\newblock Flattened-top domical water drops formed through self-organization of
  hydrophobin membranes: A structural and mechanistic study using atomic force
  microscopy.
\newblock \emph{ACS nano}, 10\penalty0 (1):\penalty0 81--87, 2016.

\end{thebibliography}

\end{document}